\documentclass[superscriptaddress,nofootinbib,12pt]{revtex4-1}
\usepackage[utf8]{inputenc}

\usepackage[dvipsnames]{xcolor}
\definecolor{red}{rgb}{0.9, 0,0}
\definecolor{cerulean}{rgb}{0., 0.42,0.9}
\definecolor{navy}{rgb}{0.05, 0.05,0.8}

\usepackage[colorlinks]{hyperref}
\hypersetup{
    colorlinks = true,
    citecolor  = red,
	linkcolor  = navy
}

\usepackage{bbm}
\usepackage{eufrak}
\usepackage{amsfonts}
\usepackage{mathrsfs}
\usepackage{bbold}
\usepackage{amsmath,amssymb}
\usepackage{subcaption}
\usepackage{siunitx}
\usepackage{slashed}
\usepackage{epsfig}
\usepackage{graphicx} 
\usepackage{url}
\usepackage{float}
\usepackage{soul}
\usepackage{pstricks}
\usepackage{color}
\usepackage{multirow}

\def\bit{\begin{itemize}}
\def\eit{\end{itemize}}
\def\ben{\begin{enumerate}}
\def\een{\end{enumerate}}

\def\vecdot#1#2{#1 \cdot #2}

\newcommand\DN[1][\relax]{%
\ifx\relax#1\relax\else{}^{#1}\fi \!X}

\newcommand{\dm}{\textrm{DM}}

\DeclareMathAlphabet\mathbfcal{OMS}{cmsy}{b}{n}

\def\sub{\text{sub}}
\def\vec#1{\mathbf{#1}}
\def\min{\text{min}}
\def\max{\text{max}}

\def\kpc{\text{ kpc}}

\def\mylog#1{\text{log}\left( #1 \right)}

\def\th{\textsuperscript{th} }
\def\fit{\text{fit}}

\begin{document}

\title{Observability of Dark Matter Substructure with Pulsar Timing Correlations}

\author{Harikrishnan Ramani}
\affiliation{Department of Physics, University of California, Berkeley, CA 94720}
\affiliation{Theoretical Physics Group, Lawrence Berkeley National Laboratory, Berkeley, CA 94720}

\author{Tanner Trickle}
\affiliation{Department of Physics, University of California, Berkeley, CA 94720}
\affiliation{Theoretical Physics Group, Lawrence Berkeley National Laboratory, Berkeley, CA 94720}
\affiliation{Walter Burke Institute for Theoretical Physics, California Institute of Technology, Pasadena, CA 91125, USA}

\author{Kathryn M. Zurek}
\affiliation{Walter Burke Institute for Theoretical Physics, California Institute of Technology, Pasadena, CA 91125, USA}

\begin{abstract} 
Dark matter substructure on small scales is currently weakly constrained, and its study may shed light on the nature of the dark matter. In this work we study the gravitational effects of dark matter substructure on measured pulsar phases in pulsar timing arrays (PTAs). Due to the stability of pulse phases observed over several years, dark matter substructure around the Earth-pulsar system can imprint discernible signatures in gravitational Doppler and Shapiro delays. We compute pulsar phase correlations induced by general dark matter substructure, and project constraints for a few models such as monochromatic primordial black holes (PBHs), and Cold Dark Matter (CDM)-like NFW subhalos. This work extends our previous analysis, which focused on static or single transiting events, to a stochastic analysis of multiple transiting events. We find that stochastic correlations, in a PTA similar to the Square Kilometer Array (SKA), are uniquely powerful to constrain subhalos as light as $\sim 10^{-13}~M_\odot$, with concentrations as low as that predicted by standard CDM.
\end{abstract}

\maketitle

\tableofcontents

\newpage

\section{Introduction}
\label{sec:introduction}

The nature of the dark matter and its associated forces, known as the dark sector, remains unknown.  To uncover its identity, interactions with the Standard Model have been probed through production at colliders, direct detection in laboratory experiments, indirect detection of dark matter annihilation products in the galaxy, and the impact of the dark sector on stellar and cosmological evolution.  However, dark matter may interact with the Standard Model only via gravity. If this is the case, gravitational probes of dark matter substructure will be the only avenue to learn more about the underlying theory of dark matter. 

The dark matter halo structure observed on cosmological and galactic scales is observed to be consistent with adiabatic density perturbations generated by inflation, and (at least at leading order) is independent of the particle nature of dark matter. On smaller mass scales, however, many theories of dark matter leave unique fingerprints on primordial density perturbations that grow into characteristic Halo Mass Functions (HMFs). The Weakly Interacting Massive Particle (WIMP), for example, features a scale-invariant adiabatic primordial power spectrum set by inflation, with a characteristic damping below $10^{-6}~M_\odot$ in the HMF, due to WIMP free-streaming \cite{Green2005a}.  On the other hand the QCD axion has large isocurvature fluctuations that can collapse to form very dense halos called miniclusters \cite{Hogan1988a,Kolb:1993zz,Kolb:1993hw,Zurek2007a,Buschmann:2019icd,Arvanitaki:2019rax}, enhancing the HMF on small scales. Many other theories predict enhanced matter power on small scales, including vector bosons produced during inflation \cite{Graham:2015rva} and theories with early matter domination \cite{Erickcek:2011us,Barenboim:2013gya,Fan:2014zua,Blinov:2019jqc}.  At present, these theories are poorly constrained by observations.

The challenge of observing structure on scales much smaller than galaxies arises because, once the virial temperature of halos halos drops below the baryon temperature, baryons no longer effectively trace the dark matter halos.  At masses below $\sim 10^9~M_\odot$, star formation is suppressed so that stars cannot be used to trace dark matter. Thus neither galaxy surveys nor observations of the Lyman-$\alpha$ absorption of the spectra of distant quasars can give information on dark matter halos on comoving scales below $\sim 0.01 - 0.1 \mbox{ Mpc}$. Smaller structures have instead been observed with strong lensing of quasars \cite{Gilman:2019nap} and with fluctuations in stellar streams \cite{Bonaca:2018fek}, both confirming subhalos down to about $10^7 ~M_\odot$. $21$~cm cosmology \cite{Munoz:2019hjh} for masses in the range $10^6-10^8 M_\odot$, strong gravitational lensing \cite{Rivero:2017mao} and stellar wakes \cite{Buschmann:2017ams} for $M > 10^5~M_\odot$, astrometric lensing \cite{VanTilburg:2018ykj, Mondino:2020rkn, Mishra-Sharma:2020ynk} for $M > 1~M_\odot$, and disruption of compact stellar systems \cite{Brandt2016a} for $M > 5 ~M_\odot$, have all been proposed to extend constraints on the HMF to lower masses.  

For sub-solar mass halos, microlensing of stars towards the Large Magellanic Cloud ({\em e.g.}, MACHO \cite{Alcock_2001}, EROS \cite{Tisserand_2007}, OGLE \cite{Wyrzykowski_2011}), Andromeda ({\em e.g.} SUBARU \cite{Niikura_2019,Smyth2019a}) or stars in the local neighborhood (from Gaia \cite{Mondino:2020rkn} and KEPLER \cite{Griest2014a}) constrains sufficiently dense halos to be a sub-dominant component of the dark matter. However, microlensing becomes ineffective in detecting subhalos below $\sim10^{-10}-10^{-11}~M_\odot$~\cite{Smyth2019a}. In the future, lensing of gamma ray bursts \cite{Katz:2018zrn} and fast radio bursts \cite{Katz:2019qug} may be able to reach these small masses, but these searches are typically only sensitive to very compact objects rather than halos. Astrometric lensing cannot constrain halos even a thousand times more dense than the local dark matter density, while micro-lensing loses reach even for halos $10^{17}$ times more dense (see for example Ref.~\cite{VanTilburg:2018ykj, Dror2019a, Croon:2020wpr}); Cold Dark Matter (CDM) halos, and even axion-like or scalar miniclusters (as discussed in Ref.~\cite{Zurek2007a}), are often too `fluffy' to be observed, particularly with microlensing. Recently, photometric monitoring of caustic transiting stars has been proposed as a probe of subhalos down to $10^{-15} M_\odot$ \cite{Dai:2019lud} and with a lower central density, although this requires dedicated monitoring by telescopes such as the Hubble Space Telescope or the James Web Space Telescope.

For low mass and low concentration subhalos, Pulsar Timing Arrays (PTAs) are a unique and powerful probe of dark matter substructure, as considered in Refs.~\cite{Siegel:2007fz, Seto:2007kj, Clark:2015sha, Schutz:2016khr, Baghram2011a, Kashiyama2012a, Kashiyama:2018gsh,Dror2019a}. We previously demonstrated that individual transiting subhalos and PBHs can be detected in the future by the Square Kilometer Array (SKA)~\cite{Rosado2015a} in the range $10^{-11}-10^3~M_\odot$~\cite{Dror2019a}. Owing to the sensitivity of an individual pulsar far exceeding that of a traditional gravitational lens, we showed in Ref.~\cite{Dror2019a} that PTA constraints on dark matter substructure remain in force (over certain mass ranges) even for halo concentration typical of ordinary CDM subhalos. This implies that they will have great sensitivity to a wide range of models with even a moderate amount of additional matter power on small scales.

To cover this wide mass range, we considered four different signal types: \textit{static} and \textit{dynamic} signals from {\em Doppler} or {\em Shapiro} effects induced by a single transiting subhalo. The Doppler delay is an acceleration effect from the subhalos gravitationally pulling the Earth or pulsars; the Shapiro delay is a gravitational redshift effect on the travel time of photons due to metric perturbations along the photon trajectory. Static and dynamic signals are differentiated by the time scale of their events.  A static signal persists over the observing time and leaves its imprint on the, usually small, second derivative of the pulsar frequency; a dynamic signal is shorter than the observing time, and gives rise to a characteristic signal shape.  In Ref.~\cite{Dror2019a} constraints in the dynamic regime were set using only the single strongest event (statistically drawn from a spatial distribution of halos). Because the characteristic signal shape is predictive, one can filter the data for the signal shape on an event-by-event basis; we will refer to single dynamic signals as {\em deterministic} throughout this work. 

The natural extension of this deterministic analysis is to study the effect from an ensemble of events, where the observable is a correlation of signal shapes and the signal is stochastic in nature. The purpose of this paper is to compute the reach on dark matter substructure using a stochastic signal for the Doppler and Shapiro delays. Such a statistical observable was considered previously in Ref.~\cite{Baghram2011a}; the present study improves over the previous analysis in important ways, by taking into account finite volume effects and the impact of the pulsar fit parameters on the signal-to-noise ratio (SNR). Additionally we provide the correct subtraction procedure to capture the effect of the pulsar model fit on a general signal. We perform this subtraction for the stochastic signal, as well as the deterministic signal of Ref.~\cite{Dror2019a}; for the latter we find the pulsar model fit gives rise to a substantial correction on the reach. Note that the actual impact of the pulsar model fit will be dependent on the precise timing model and the pulsar sample, and we leave an analysis utilizing existing data for future work.

The outline of this paper is as follows. We begin by determining how a stochastic dark matter signal affects the PTA observable \textit{i.e.} the residual phase, in Sec.~\ref{sec:signal_analysis}.  This includes a general discussion of the pulsar model fit and dark matter signal in Sec.~\ref{subsec:pulsar_phase_corr}, a concrete calculation of the signal correlator (for both Doppler and Shapiro delays) in Sec.~\ref{subsec:compute_signal_corr}, and a derivation of the optimal signal-to-noise ratio (SNR) in Sec.~\ref{subsec:signalSNR}. We then turn to dark matter model reach in Sec.~\ref{sec:results}. We compute constraints for monochromatic mass distribution of PBHs and more diffuse halos in Sec.~\ref{subsec:mono_mass_dist}, and generalize these results in Sec.~\ref{sec:ext_mass_dist} to a slightly broadened HMF, along with a CDM-like HMF. Lastly we conclude with future directions for applying our results to existing PTA data, and to a broader class of dark matter models. 

\section{Dark Matter Signatures in Pulsar Phase Correlations}
\label{sec:signal_analysis}

The goal of this section is to compute the signal-to-noise ratio (SNR) in PTAs generated by an ensemble of transiting dark matter subhalos. We begin with a discussion of the PTA observable, pulsar phases, and how dark matter subhalos can produce correlations in them. These correlations are the signal which we compare with PTA timing noise to construct an SNR. This SNR, for a general model of dark matter substructure, will then be our basis for projecting constraints in the next section.

\subsection{Pulsar Phase Correlator}
\label{subsec:pulsar_phase_corr}

Pulsars with millisecond periods, observed over decades, are known to be good clocks. This is because, while the pulsar period may fluctuate on short time scales, these fluctuations do not accumulate, such that the arrival time of light pulses can be predicted with a simple model of the pulsar phase evolution, 
\begin{equation}
\phi(t) = \phi^0 + \nu t + \frac{1}{2}\dot{\nu} t^2,
\label{eq:timingmodel}
\end{equation} 
where $\phi^0, \nu, \dot{\nu}$, are the phase offset, pulsar frequency, and its first time derivative. The success of this model implies that any deviations due to dark matter substructure can be observed or constrained. These deviations are characterized by the residual phase, 
\begin{align}
s(t) \equiv \phi(t) - \phi_\fit(t),
\end{align}
where $\phi_\fit = \phi^0_\fit + \nu_\fit t + \dot{\nu}_\fit t^2/2$, and $\phi^0_\fit, \nu_\fit, \dot{\nu}_\fit$ are obtained by fitting the measured pulsar phase with the timing model. In the absence of dark matter substructure, this residual is well-fit by stationary white noise, $s(t) = n(t)$, where $n(t)$ is defined by its statistical properties, $\langle n(t) n(t') \rangle = \nu^2 t_\text{rms}^2 \Delta t \, \delta (t - t')$, with $\Delta t$ the measurement cadence and $t_\text{rms}$ the root-mean-square post-fit timing residual, discussed further in Appendix~\ref{app:SNR_derivation}. 

If dark matter substructure is present, the residual phase will have additional contributions, which we quantify as 
\begin{equation}
s(t) = h(t) + n(t) \, ,
\end{equation}
with $h(t)$ the {\em subtracted} dark matter signal,
\begin{equation}
h(t) \equiv \delta \phi(t) - \delta {\phi}_\fit(t) \, . 
\label{eq:subtracted_signal}
\end{equation}
Here $\delta \phi$ is the phase modification induced by the dark matter substructure, and ${\delta \phi}_\fit$ is the part of the signal absorbed by the pulsar timing model fit, as detailed in Appendix~\ref{app:subtraction_procedure}. $\delta \phi$ can be written in terms of a frequency shift, 
\begin{equation}
\delta \phi(t) = \int_0^t \delta \nu(t') \,  dt' \, .
\label{eq:delta_phi_delta_nu}
\end{equation}

We consider two gravitational effects from transiting subhalos that induce a frequency shift. The Doppler effect arises when transiting subhalos induce an acceleration in the Earth or pulsar, while the Shapiro effect is due to the change in the gravitational potential along the photon's trajectory; see Refs.~\cite{Siegel2007a,Seto2007a,Baghram2011a,Kashiyama2012a,Dror2019a} for more details. These shifts, for a single transiting subhalo, are given by
\begin{align}
\left(\frac{\delta \nu}{\nu}\right)_D & =\, \mathbf{\hat{d}} \cdot \int \nabla \Phi(\vec{r}, M) \;dt \label{eq:delta_nu_D}\\
\left(\frac{\delta \nu}{\nu}\right)_S & = -2 \int \mathbf{v} \cdot \nabla \Phi(\vec{r}, M) \;dz \, \label{eq:delta_nu_S} ,
\end{align}
for the Doppler and Shapiro delay respectively, where $\Phi$ is the gravitational potential from a single subhalo, $\vec{v}$ is the subhalos velocity, and $\vec{\hat{d}}$ is the direction from the Earth to the pulsar. In Ref.~\cite{Dror2019a} these expressions were utilized to constrain the abundance of PBHs and compact halos via {\em single} transiting subhalos, though only the signal from the closest subhalo was considered. The advantage of this approach is the ability to predict the specific signal shape in order to filter the data accordingly. The disadvantage is that, at small masses, even the closest subhalo does not produce a measurable signal. However, in this small mass regime there is an abundance of subhalos which could cumulatively leave a discernible signal. This signal from a {\em statistical ensemble} of transiting subhalos,
\begin{align}
\delta \phi(t) & = \sum_{i = 1}^N \delta \phi_i(t) \, ,
\label{eq:statphaseshift}
\end{align}
where $ \delta \phi_i$ is the phase modification from the $i$\th event, is our starting point. Correlations can then be written, 
\begin{align}
\langle \delta \phi(t) \delta \phi(t') \rangle & = \sum_{i = 1}^N \langle \delta \phi_i(t) \delta \phi_i(t') \rangle + \sum_{i \neq j}^{N(N-1)} \langle \delta \phi_i(t) \delta \phi_j(t') \rangle \equiv R_1(t, t') + R_2(t, t') \, , \label{eq:signal_corr}
\end{align}
where $R_1$ ($R_2$) contains contributions from averaging over one (two) subhalo(s). 

\subsection{Dark Matter Signal Correlator}
\label{subsec:compute_signal_corr}

We now compute the dark matter induced phase correlation, $R$ in Eq.~\eqref{eq:signal_corr}, for the Doppler and Shapiro delays. The expectation, $\langle \rangle$, averages over the random variables that determine the phase shifts from all $N$ subhalos. Similar to the treatment in Ref.~\cite{Rivero:2017mao}, we take these random variables to be the subhalo masses, $M_i$, initial positions $\vec{r}^0_i$, and velocities $\vec{v}_i$. The 1-subhalo term, $R_1$, in $R$ does not include subhalo correlations, while $R_2$ depends on the subhalo correlation power spectrum, $P_\xi$, defined from the subhalo number density, $n_\text{sub}$ as,
\begin{align}
n_\sub(\vec{x}) & \equiv \overline{n} \left(1 + \delta_n(\vec{x}) \right) \\
\langle \delta_n \rangle_\text{en} & = 0 \\
\langle \delta_n(\vec{x}) \delta_n(\vec{y}) \rangle_\text{en} & \equiv \xi(\vec{x} - \vec{y}) \\
P_\xi(\vec{k}) & = \int d^3\vec{x} \, e^{i \vec{k} \cdot \vec{x}} \xi(\vec{x})\label{eq:position_matter_power_relationship} \, .
\end{align}
where $\langle \rangle_\text{en}$ denotes an ensemble average.\footnote{For a monochromatic mass distribution, one can show that the matter power spectrum, $P_m$, and the correlation power spectrum, $P_\xi$, are related by $P_m = |W(k, M)|^2/\overline{n} + P_\xi \, |W(k, M)|^2$} The statistical nature of the signal is similar to Ref.~\cite{Baghram2011a}, with important differences accounting for finite observation volume.  The comparisons between the formalisms is discussed in Appendix~\ref{powerallspace}. We leave a discussion of $R_2$, which is non-zero when $P_\xi \neq 0$, for future work, as we expect $R_2$ to be subdominant to the leading effect from uniformly distributed subhalos. For notational simplicity, we will refer to $R_1$ as $R$ from here on.  

Assuming $M_i, \vec{r}^0_i, \vec{v}_i$ are independent and have identical probability distribution functions, we can write $R$ in terms of the frequency shift, averaged over a single subhalo of mass $M$ with position ${\bf r}_0$ and velocity ${\bf v}$,
\begin{align}
R(t, t') & = \int_0^t \int_0^{t'} dt_1 dt_2 \int d^3\vec{v} f_\vec{v}(\vec{v}) \int \frac{dM}{M} F(M) \int \, d^3\vec{r}^0  \, \delta \nu(t_1; M, \vec{r}^0, \vec{v}) \delta \nu(t_2;, M, \vec{r}^0, \vec{v}) \, ,
\label{eq:R1_correlator}
\end{align}
where $F(M) = dn/d\log{M}$ is the HMF, $f_\vec{v}$ is a boosted Maxwell-Boltzmann velocity distribution, with $v_0 = 230$ km$/$s, $v_E = 240$ km$/$s, and $v_\text{esc} = 600$ km$/$s. Assuming uniformly distributed subhalos means the probability distribution function in $\vec{r}^0$ is simply $1/V$, where $V$ is the observing volume. We expect a relatively weak dependence on the velocity distribution, and from here on will take the velocity to be an average value. The Doppler delay signal will depend on the average velocity, $\langle v \rangle_\vec{v} \equiv \overline{v} \approx 340 $ km$/$s, whereas the Shapiro delay will depend on the velocity component perpendicular to the Earth pulsar direction, $\langle v_\perp \rangle_\vec{v} \equiv \overline{v}_\perp \approx 270$ km$/$s.\footnote{To compute $\overline{v}_\perp$ we calculate the expectation value of $\sqrt{\vec{v} - ( \vec{v} \cdot \vec{\hat{d}} )\vec{\hat{d}}}$ (the magnitude of the components perpendicular to the Earth-pulsar direction, $\vec{\hat{d}}$) and average over all directions of the Earth velocity, $\vec{\hat{v}}_E$}

We now compute $R$ for the Doppler and Shapiro delays. 

\subsubsection{Shapiro delay}
\label{sec:compute_signal_corr_shap}

We begin by simplifying the building block of a statistical signal by writing the Shapiro delay from a single subhalo, $\delta \nu_S$ in Eq.~\eqref{eq:delta_nu_S}, as
\begin{align}
\delta \nu_S & = -8 \pi i \nu G M  \int_0^{z_0} d\ell \int \frac{d^3k}{(2 \pi)^3} \frac{\vec{v} \cdot \vec{k}}{k^2} W(k, M) e^{i k_z (\ell - z)} e^{-i \vecdot{\vec{k}_\perp}{\vec{r}_\perp}} \, ,
\end{align}
where we have used the relation between the gravitational potential $\tilde \Phi$, density profile $\tilde{\rho}(k,M)$, and window function $W(k,M)$: $\widetilde{\Phi}(k) = -(4 \pi G/k^2) \tilde{\rho}(k,M) \equiv -(4 \pi G M/k^2) W(k, M)$.  It will be useful to define a coordinate system with $\vec{\hat{z}}$ along the Earth-pulsar direction, and a plane perpendicular to $\vec{\hat{z}}$ such that $\vec{r}_\perp = \vec{v}_\perp t + \vec{r}^0_\perp = \vec{v}_\perp(t - t^0) + \vec{b}$, with $\vec{b}$ the impact parameter.  The position of a subhalo is thus $\vec{r} = \vec{r}_\perp + \hat{\vec{z}} (\vec{r} \cdot \hat{\vec{z}})$.   

Evaluating the $k_z$, $z$ integrals, in the limit $k_z \lesssim z_0^{-1} \ll k_\perp$ and $0 < z < z_0$, gives (similar to Ref.~\cite{Baghram2011a})
\begin{align}
\delta \nu_S  \approx 8 \pi i \nu G M  \int \frac{d^2k_\perp}{(2 \pi)^2} \left( \vec{v}_\perp \cdot \vec{k}_\perp \right) \frac{W(k_\perp, M)}{k_\perp^2} e^{i \vec{k}_{\perp} \cdot \vec{r}_\perp } \, .
\end{align}
This expression can be written in terms of Bessel functions
\begin{align}
\delta \nu_S   \approx - 4 G M \nu \frac{\vecdot{\vec{v}_\perp}{\vec{r}_\perp}}{r_\perp^2} r_\perp \int_0^\infty dk \, W(k, M) J_1(k \, r_\perp) \equiv - 4 G M \nu \frac{\vecdot{\vec{v}_\perp}{\vec{r}_\perp}}{r_\perp^2} \mathcal{F}(M, r_\perp) \, .
	\label{eq:sh_single_signal}
\end{align} 
We have defined a form factor ${\cal F}$,
\begin{align}
\mathcal{F}(M, x) \equiv x \int_0^\infty W(k, M) J_1(k x) \, dk \, , \label{eq:F_int}
\end{align}
with $J_1(x)$  the first order Bessel function. In the PBH limit, $\mathcal{F} = 1$, Eq.~\eqref{eq:sh_single_signal} reduces to the corresponding expression in Ref.~\cite{Dror2019a}. We further take $W(y/r_\perp, M) \approx W(y/b, M)$, which we expect to be reasonable as the signal is peaked near $t = t^0$. We then obtain
\begin{align}
\delta \nu_S &  \approx  - 4 G M \nu \, \overline{v}_\perp^2  \, \frac{t - t^0}{b^2 + \overline{v}_\perp^2 \left( t - t^0 \right)^2} \, \mathcal{F}(M, b) \, .
\label{eq:shapiro_signal_shape}
\end{align}
We can now compute the signal correlator in Eq.~\eqref{eq:R1_correlator}. In order to account for a finite observing volume we decompose the integral over the initial position as $d^3\vec{r}^0 = \overline{v}_\perp dz^0 \, db \, dt^0$, where $\vec{\hat{z}}, \vec{\hat{b}}, \vec{\hat{v}}_\perp$ are orthogonal directions. We note that in the large $t^0$ limit, $(\delta \nu / \nu)^2 \propto t_0^{-2}$, and for simplicity will extend the bounds on the $t^0$ integral to infinity. Finally,
\begin{align}
R_S(t, t') = 32 \pi G^2 \nu^2 z_0 \int dM \, M F(M) \int db \, b \, A\left(t, t', \frac{b}{\overline{v}_\perp} \right)  \mathcal{F}(M, b)^2 \, ,
\label{eq:R_sh_final}
\end{align}
where
\begin{align}
A(t, t', \tau) \equiv \int_0^t \int_0^{t'} dt_1 dt_2 \, \frac{1}{4 \tau^2 + \left( t_1 - t_2 \right)^2} \, .
\end{align}

\subsubsection{Doppler delay} 
\label{sec:compute_signal_corr_dop}

The derivation of the signal correlator for the Doppler delay begins analogously to the Shapiro delay. We write the frequency shift of an individual subhalo from the $J$\th pulsar, $\delta \nu_{D, J}$,
\begin{align}
\delta \nu_{D}^{\, J} = 4 \pi i \nu^J  G M \int_{-\infty}^t dt' \int \frac{d^3k}{(2 \pi)^3} \frac{\hat{\vec{d}}^J \cdot \vec{k}}{k^2} W(k, M) e^{-i \vec{k} \cdot \vec{r}(t')} \, ,
\end{align}
where we have again used the relation between the gravitational potential $\tilde \Phi$, density profile $\tilde{\rho}(k,M)$, and the window function $W(k,M)$: $\widetilde{\Phi}(k) = -(4 \pi G/k^2) \tilde{\rho}(k,M) \equiv -(4 \pi G M/k^2) W(k, M)$. $\hat{\vec{d}}^J$ is the the direction pointing from the Earth to the $J$\th pulsar, and $\vec{r}(t) = \vec{v}(t - t^0) + \vec{b}$, where $\vec{b}$ is the impact parameter and orthogonal to $\vec{v}$. We evaluate the $t'$ integral with the identity $\int_{-\infty}^t dt' e^{-i {\bf k} \cdot {\bf v} t'} = \frac{i}{{\bf k} \cdot {\bf v}} e^{i {\bf k} \cdot {\bf v} t} + \pi \delta ({\bf k} \cdot {\bf v})$ and decompose $\vec{k}, \vec{\hat{d}}^J$ into the coordinate system spanned by $\mathbf{b}, \mathbf{v}$, \textit{i.e.} $\vec{k} = \vec{k}_b + k_v \vec{\hat{v}}$, where $\vec{k}_b$ lies in the plane of the impact parameter (and therefore $\vec{k}_b \cdot \vec{v} = 0$).  We obtain
\begin{align}
\delta \nu_{D, J}  & = - \frac{4 \pi G M \nu^J}{\overline{v}}\int \frac{d^2k_b dk_v}{(2 \pi)^3} \frac{\mathbf{\hat{d}}^J \cdot \mathbf{k}_b}{k^2} \frac{W(k, M)}{k_v} e^{-i \mathbf{k}_b \cdot \mathbf{b}} e^{-i k_v \overline{v} (t - t^0)} \nonumber \\
&+ \frac{4 \pi^2 i G M \nu^J}{\overline{v}} \int \frac{d^2k_b}{(2 \pi)^3} \frac{\mathbf{\hat{d}}^J \cdot \mathbf{k}_b}{k_b^2} W(k_b, M) e^{-i \mathbf{k}_b \cdot \mathbf{b}} \nonumber \\
& -\frac{4 \pi G M \nu^J \, \vec{\hat{d}}^J \cdot \vec{\hat{v}}}{\overline{v}}\int \frac{d^2k_b dk_v}{(2 \pi)^3} \frac{W(k, M)}{k^2} e^{-i \mathbf{k}_b \cdot \mathbf{b}} e^{-i k_v \overline{v} (t - t^0)} \, .
\label{eq:dop_signal_no_approx}
\end{align}
The first two terms can be combined and simplified when $k_v \ll k_b$ ($\mathbf{k} \approx \mathbf{k}_b$) using the identity $\int \frac{dk_v}{2\pi} \frac{e^{-i k_v v(t-t_0)}}{k_v} = -\frac{i}{2} \frac{t - t^0}{|t - t^0|}$, and written as,
\begin{align}
\delta \nu_D^J  & = 2 \, \Theta(t - t^0) \left( \vec{\hat{d}}^J \cdot \vec{\hat{b}} \right) \frac{G M \nu^J }{b \overline{v}}  \mathcal{F}(M,b) \, , \label{eq:dop_signal_simple}
\end{align}
where $\mathcal{F}$ is defined in Eq.~\eqref{eq:F_int}.  Each event causes a jump in the pulsar frequency, signaling that a series of such step functions causes the pulsar frequency to undergo a random walk.\footnote{Ref.~\cite{Baghram2011a} dropped the second term in Eq.~\eqref{eq:dop_signal_no_approx}, which is necessary for obtaining this random walk behavior.} The third term in Eq.~\eqref{eq:dop_signal_no_approx} corresponds to a transient -- the signal does not accumulate in the pulsar frequency with each passing event --  as in Ref.~\cite{Dror2019a}; we expect such terms to have a subdominant effect and hence drop them.

For the Shapiro correlator, the signal accumulates along an Earth-pulsar path, implying that different Earth-pulsar contributions are uncorrelated. By contrast, for the Doppler correlator, a subhalo may give rise to an acceleration of the Earth alone, indicating that one event will leave a signal across the entire array of pulsars. This allows for correlation across pulsars, $\langle \delta \phi^I \delta \phi^J \rangle \equiv R^{\, IJ}$, which will result in a larger SNR than contributions without this correlation.

We substitute Eq.~\eqref{eq:dop_signal_simple} in Eq.~\eqref{eq:R1_correlator}, identifying $d^3\vec{r}^0 = \overline{v} \, d^2\vec{b} \, dt^0$. After simplification, 
\begin{align}
R_D^{\, IJ}(t, t') = \frac{4\pi G^2 \nu^2}{\overline{v}} \left( \vec{\hat{d}}^I \cdot \vec{\hat{d}}^J \right) B(t, t') \int dM \, M F(M) \int \frac{db}{b}\, \left( \mathcal{F}(M, b) \right)^2 \, ,
\label{eq:R_dop_final}
\end{align}
where the correlator for a random walk process is proportional to the minimum of the time of two events:
\begin{align}
B(t, t') \equiv \int_0^t \int_0^{t'} \text{min}(t_1, t_2) \, dt_1 dt_2 = \frac{\text{min}(t, t')^2\left( 3 \, \text{max}(t, t') - \text{min}(t,t') \right)}{6} \, .
\end{align}

\subsection{Constructing the Signal to Noise Ratio}
\label{subsec:signalSNR}

We now have all the ingredients to compute the SNR, which gives the significance of the measured dark matter signal over the pulsar timing noise. The dark matter signal is subtracted to take into account the effect of the pulsar fit model, $h = \delta \phi - \delta \phi_\fit$, as discussed in Sec.~\ref{subsec:pulsar_phase_corr} and derived in Appendix~\ref{app:subtraction_procedure}. We quote the result here,
\begin{align}
h(t) = \delta \phi(t) -\sum_{n=0}^2 \left[ \frac{1}{T} \int_0^T dt' \,  \delta \phi(t') f_n(t') \right] f_n(t) \, ,
\label{eq:subtracted_single_signal_expanded}
\end{align}
where $f_n(t) = \sqrt{2n + 1} P_n(2t/T - 1)$, and $P_n$ are the Legendre polynomials. The sum is from zero to two in order to include $\phi^0,~\nu,$ and $ \dot{\nu}$ in the timing model fit. It follows that correlators of $h$, $R_\text{sub}(t,t') \equiv \langle h(t) h(t') \rangle$, are related to correlators of $\delta \phi$ and $R$ by (see details in Appendix~\ref{app:subtraction_procedure}):
\begin{align}
R_{\rm sub}(t, t') &= R(t, t') - \sum_{n = 0}^2 f_n(t) \mathcal{R}_n(t') - \sum_{n = 0}^2 f_n(t') \mathcal{R}_n(t) + \sum_{n=0}^2 \sum_{m=0}^2 \mathcal{R}_{nm} f_n(t) f_m(t') \\
\mathcal{R}_n(t') & \equiv \frac{1}{T} \int_0^T R(t, t') f_n(t) \, dt \\
\mathcal{R}_{nm} & \equiv \frac{1}{T^2} \int_0^T \int_0^T dt dt' \, R(t, t') \,  f_n(t) f_m(t') \, .
\label{eq:sub_auto_corr_main}
\end{align}

Having defined the subtracted signal we construct the SNR using a matched filter procedure as in Refs.~\cite{Moore2015a,Smith2019a}. We begin with the deterministic signal, as in Ref.~\cite{Dror2019a}, where we study the SNR, $\text{SNR}_\text{det}$, from the subhalo which imprints the largest signal. The expressions for the SNR are derived in Appendix~\ref{subapp:deterministic_snr}, and depend on whether the signal is uncorrelated across pulsars (the `pulsar' term, $\text{SNR}_{\text{det},P}$) or correlated (the `Earth' term, $\text{SNR}_{\text{det},E}$):
\begin{align}
\text{SNR}_{\text{det}, P}^2 & = \frac{1}{\widetilde{N}} \, \underset{\{I\}}{\max} \left[ \int_0^T dt \,  h_I^2(t) \right] \label{eq:snr_det_pulsar}\\
\text{SNR}_{\text{det}, E}^2 & = \frac{N_P}{\widetilde{N}} \int_0^T dt \, \langle h^2(t) \rangle_\mathcal{P} \label{eq:snr_det_Earth},
\end{align}
where the maximum over $I$ denotes the maximum signal across all pulsars in the array. $\langle \rangle_\mathcal{P}$ denotes averaging over the pulsar positions, $\widetilde{N} = \nu^2 t_\text{rms}^2 \Delta t$, $T$ is the observing time, $\Delta t$ the cadence, and the residual timing noise is $t_\text{rms}$. 

There are two key differences between the SNR here and in Ref.~\cite{Dror2019a}. The first is that, in order to unify the formalism presented here with the analysis in Ref.~\cite{Dror2019a}, we cast the signal in the residual phase and not residual frequency shift, $\delta \nu$ (related to each other by Eq.~\eqref{eq:delta_phi_delta_nu}). Second, and more importantly, the signal in the SNR is the subtracted signal, $h$, as opposed to $\delta \phi$, which was neglected in Ref.~\cite{Dror2019a}. This causes an $\mathcal{O}(1)$ difference in the overall SNR and is discussed in more detail in Sec.~\ref{subsec:mono_mass_dist}.

The pulsar and Earth term SNR for the stochastic signal ($\text{SNR}_P$ and $\text{SNR}_E$ respectively) are derived in Appendices~\ref{subapp:stochastic_snr_pulsar},~\ref{subapp:stochastic_snr_Earth}, and we quote the result here, 
\begin{align}
\text{SNR}_{P}^2 & = \frac{N_P}{2\widetilde{N}^2} \int dt dt' \langle R^\text{sub}_I(t, t')^2 \rangle_{\mathcal{P}} \label{eq:SNR_P_simple} \\
\text{SNR}_{E}^2 & = \frac{N_P(N_P - 1)}{2 \widetilde{N}^2} \int dt dt' \left\langle R^\text{sub}_{IJ}(t, t')^2 \right\rangle_{\mathcal{P}} \, .\label{eq:SNR_E_simple}
\end{align}
The indices $I,J$ run over the pulsars in the array and $\langle \rangle_{\mathcal{P}}$ averages over the pulsar positions. The pulsar timing array parameters, $N_P$, $T,~\Delta t,~t_{\rm rms}$ are drawn based on the capabilities of current PTAs, extrapolated to the potential of future PTAs.  The currently operating PTAs are European Pulsar Timing Array (EPTA)~\cite{Desvignes2016a}, Parkes Pulsar Timing Array (PPTA)~\cite{Manchester2013b}, North American Nanohertz Observatory for Gravitational Waves (NANOGrav)~\cite{Arzoumanian2016a}; the MeerKAT telescope has a pulsar timing program (MeerTime~\cite{Bailes2018a}), along with the Five-hundred-meter Aperture Spherical Telescope (FAST)~\cite{Hobbs2019a}. The International Pulsar Timing Array (IPTA)~\cite{Perera2019a} is comprised of EPTA~\cite{Desvignes2016a}, PPTA~\cite{Reardon2015a}, and NANOGrav~\cite{Arzoumanian2015a}, and between the three collaborations has measured $N_P = 73$ unique millisecond pulsars for $T \sim 10-30$ years, with timing residuals in the range $t_\text{rms} =50 ~\textrm{ns}-10^4~\textrm{ns}$, at a distance of $z_0 \sim 1-5$ kpc, and a cadence of $\Delta t \sim 1-4$ week. The future Square Kilometer Array (SKA)~\cite{Rosado2015a} could  increase the number of pulsars to $N_P = 200$, with timing residuals of $t_\text{rms} \sim 50$ ns. The FAST telescope could optimistically reduce the timing residuals to $t_\text{rms} \sim 1-10$ ns, assuming the current limitation is statistics~\cite{Hobbs2019a}. Our baseline PTA parameters, which we utilize in the next section, and assume when simplifying analytic results, are based on the estimated capability of the future SKA PTA: $N_P = 200$, $t_\text{rms} = 50 $ ns, $\Delta t = 2$ week, $T = 20$ years, $z_0 = 5$ kpc.

\section{Observability of Dark Matter Substructure with Pulsar Timing Arrays}
\label{sec:results}

We can now determine the observability of dark matter substructure.  In this Section, we consider both monochromatic and CDM-like HMFs, with constituent subhalos having either PBH or NFW density profiles. We determine the constraints on the mass fraction, $f \equiv \Omega/\Omega_\text{DM}$, in these models with a future PTA with SKA-like capabilities (defined in Sec.~\ref{subsec:signalSNR}).

\subsection{Monochromatic Mass Distribution}
\label{subsec:mono_mass_dist}

\begin{figure}[t]
\includegraphics[width=0.475\textwidth]{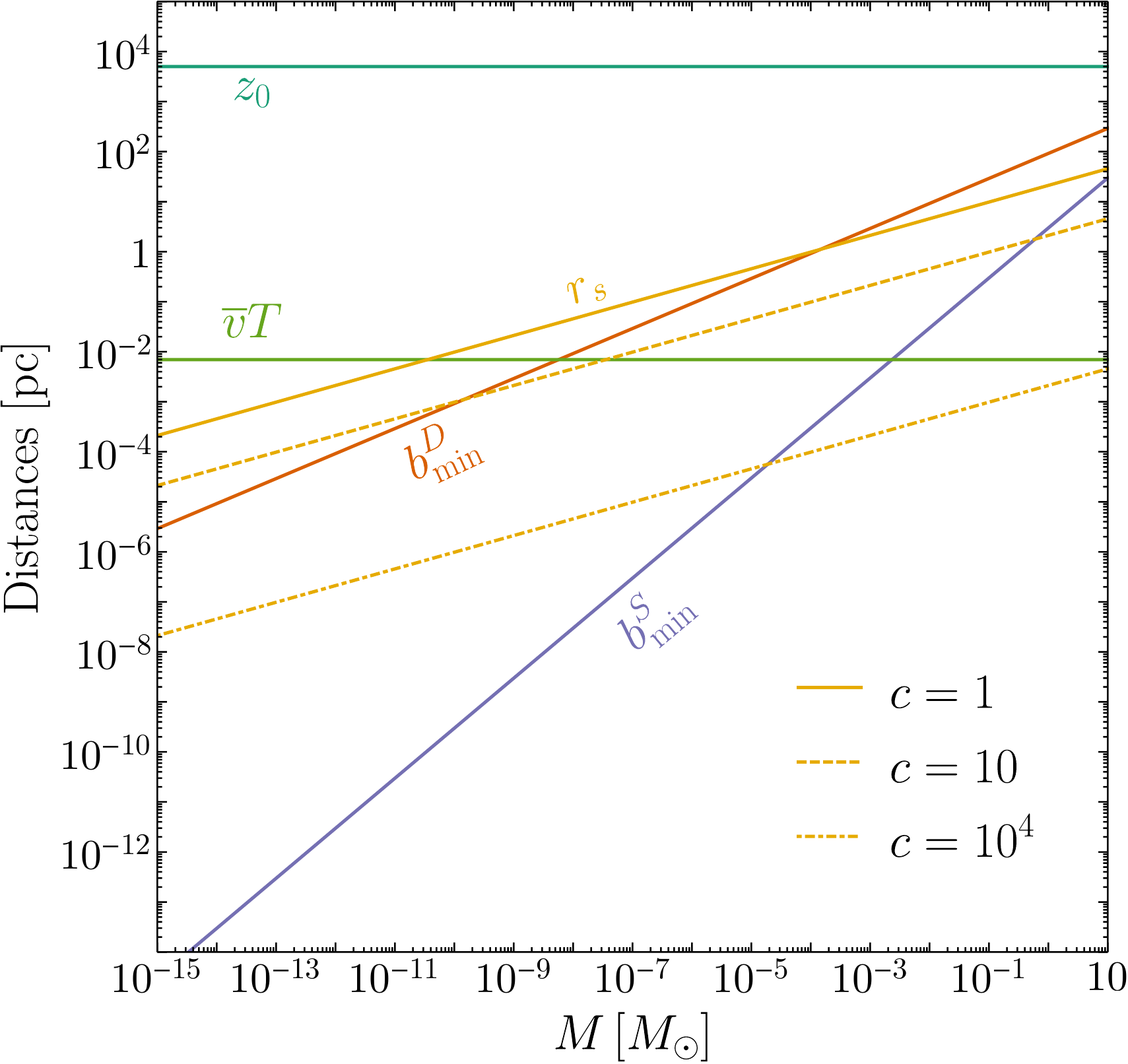}
\caption{Relevant length scales for measurement of dark matter subhalos with PTAs, as a function of the subhalo mass $M$ (assuming $f = 1$). $z_0$, the largest scale, is the Earth-pulsar distance.  $b_\text{min}^{D(S)} $ denote subhalo impact parameters for the Doppler (Shapiro) searches (see text for details); to have sensitivity they must be smaller than $\overline{v}_{(\perp)} T$, showing why the Doppler search is sensitive to lower mass subhalos than the Shapiro search. The yellow lines are the scale radius, $r_s$, for $c = 1, 10, 10^4$. A subhalo with $r_s > b_\text{min}$ has weakened constraints, and when $r_s > v T$ the constraints are negligible, indicating the subhalo concentration parameters to which PTAs are sensitive.}
\label{fig:relevant_distances}
\end{figure}

To gain intuition for how PTAs derive the power of their constraints, we begin with the simplest case of a monochromatic mass distribution of subhalos of mass $M$, $F(M') = \overline{n} M' \delta(M - M')$.  Before deriving these constraints in detail it will be important to understand the length scales which shape our results, summarized in Fig.~(\ref{fig:relevant_distances}).
The first important length scales are the range of impact parameters of transiting subhalos to which the PTA measurement is sensitive.  The distance a subhalo can travel over the observation time $T$ is  
\begin{align}
b_\text{max} \sim 10^{-8} \text{ Mpc} \left( \frac{v}{10^{-3} \, c} \right) \left( \frac{T}{20 \text{ yr}} \right) \, ,
\label{eq:b_max}
\end{align}
shown in Fig.~(\ref{fig:relevant_distances}). A subhalo must have an impact parameter smaller than $v T$ to be observable in a stochastic or dynamic search. On the lower end of the impact parameter range, the smallest impact parameter over an ensemble of $N$ events can be derived (from evenly spatially distributed subhalos), as in Ref.~\cite{Dror2019a}, for the Shapiro and Doppler delays,
\begin{align}
b_\min^S & \sim \frac{\overline{v}_\perp T}{N} \quad , \quad b_\min^D \sim  \frac{3}{2} \frac{\overline{v} T}{\sqrt{N}} \, ,
\label{eq:b_min_statistics}
\end{align}
where we have quoted the 90\th percentile result. The number of subhalos in the observing volume is $N = f \rho_\dm V/M$. The observing volume (as appears in Eq.~\eqref{eq:R1_correlator}) is a cylinder for the Doppler effect, $V_D = \pi \overline{v}^3 T^3$, and a rectangular box for the Shapiro effect, $V_S = \overline{v}_\perp^2 T^2 z_0$. We have the minimum impact parameters labeled $b_{\rm min}^S,~b_{\rm min}^D$ in Fig.~(\ref{fig:relevant_distances}) for $f = 1$. In order to constrain a subhalo of mass $M$, this minimum impact parameter must be less than the maximum in Eq.~\eqref{eq:b_max}, and by examining Fig.~(\ref{fig:relevant_distances}), we see that a Shapiro search will have greater sensitivity to larger mass subhalos than a Doppler signal.

For subhalos with NFW profile the size of the subhalo is also a relevant scale. There are two important sizes, the virial radius, $r_v$, which contains all the mass, and the scale radius, $r_s$,
\begin{align}
r_s\sim \frac{10^{-5} \text{ Mpc}}{c} \left( \frac{M}{M_\odot} \right)^\frac{1}{3} \, ,
\label{eq:rs}
\end{align}
which quantifies the compactness of a subhalo, parameterized by the concentration parameter, $c \equiv r_v/r_s$. If $b > r_v$ then the subhalo can be treated as point-like, and if $b > r_s$, there is only a modest loss in sensitivity (as we will discuss in more detail below, Sec.~(\ref{sec:mono_mass_dist_co})).  From Fig.~(\ref{fig:relevant_distances}), we see that the Doppler search in particular has strong sensitivity to low concentration subhalos, while the Shapiro search will more rapidly lose its reach for extended subhalos in comparison to PBHs. 

We now discuss constraints for point-like (PBH) and NFW density profiles in more detail.

\subsubsection{Point-like subhalo (PBH)}
\label{sec:mono_mass_dist_pbh}

\begin{figure}[ht]
\includegraphics[width=\textwidth]{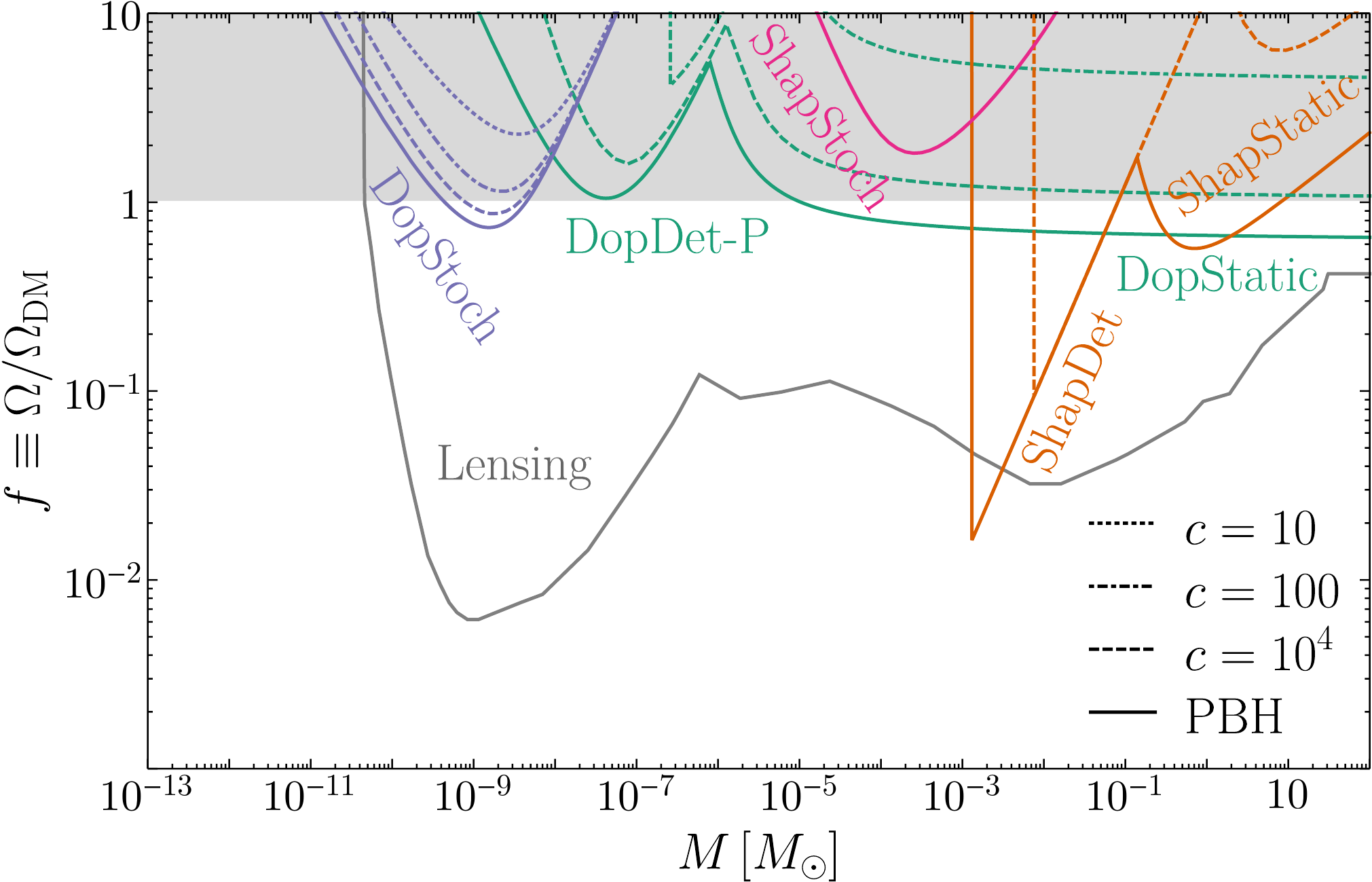}
\caption{Limits from PTAs on the dark matter mass fraction $f = \Omega / \Omega_\text{DM}$ in subhalos of mass $M$ for different subhalo concentration parameters, $c = 10, 100, 10^4$, and the PBH limit, $c \rightarrow \infty$. Results derived in Ref.~\cite{Dror2019a} from deterministic single transiting objects and static signals are labeled `DopDet-P', `DopStatic', `ShapDet', and `ShapStatic' and shown in green and orange. The `DopDet-P' and `ShapDet' constraints have been weakened relative to Ref.~\cite{Dror2019a} due to the subtraction procedure discussed in Appendix~\ref{app:subtraction_procedure}. New results of this paper utilizing a stochastic signal induced by multiple transiting subhalos are labeled `DopStoch' and `ShapStoch', and shown in blue and pink, respectively. An SKA-like PTA, described in Sec.~\ref{subsec:signalSNR}, with identical pulsars was assumed. Lensing constraints in gray are from Refs.~\cite{Alcock_2001, Tisserand_2007, Wyrzykowski_2011, Zumalacarregui2018a, Niikura_2019,Smyth2019a}, and disappear for $c < 10^7$.}
\label{fig:monochromatic_mass_constraints}
\end{figure}

\begin{figure}[ht]
\includegraphics[width=\textwidth]{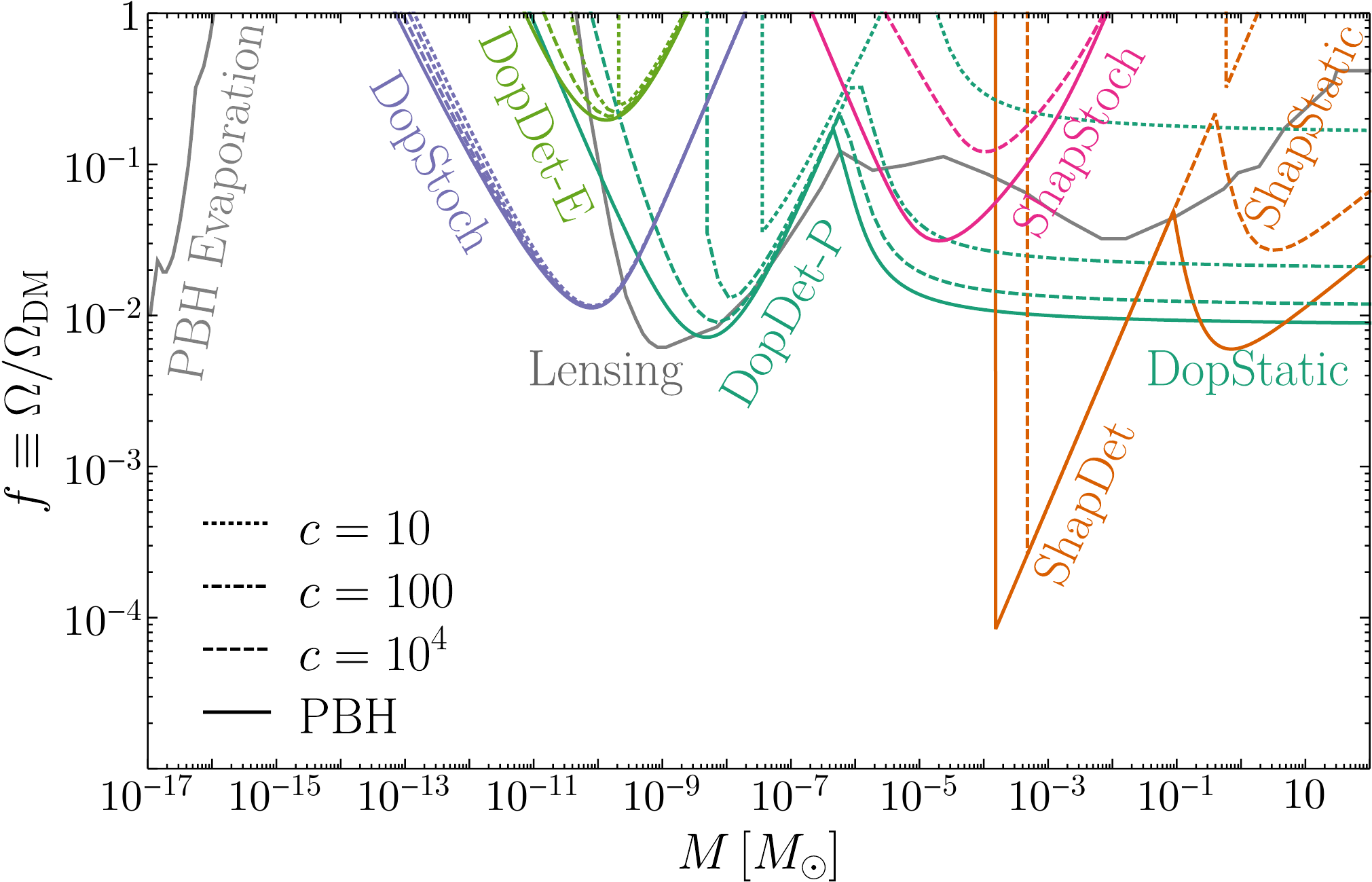}
\caption{Similar to Fig.~(\ref{fig:monochromatic_mass_constraints}) we show limits from PTAs on the dark matter mass fraction $f = \Omega / \Omega_\text{DM}$ for $c = 10, 100, 10^4$ and the PBH limit.  Here we have assumed a more optimistic set of PTA parameters compared to Fig.~(\ref{fig:monochromatic_mass_constraints}): $N_P = 1000$, $T = 30 \text{ yr}$, $t_\text{rms} = 10 \text{ ns}$, $\Delta t = 1 \text{ week}$, and $z_0 = 10 \text{ kpc}$. One additional constraint `DopDet-E' is visible here and corresponds to the deterministic Doppler Earth term. Lensing constraints are again from Refs.~\cite{Alcock_2001,Tisserand_2007,Wyrzykowski_2011,Zumalacarregui2018a,Niikura_2019,Smyth2019a}, and disappear for $c < 10^7$. The PBH evaporation constraint (`PBH Evaporation') is from Ref.~\cite{Laha:2020ivk}.}
\label{fig:monochromatic_mass_constraints_opt}
\end{figure}

We begin by laying out the constraints in the simplest case: a point-like subhalo (or PBH), where the impact form factor in Eq.~\eqref{eq:F_int} simplifies to $\mathcal{F} = 1$. The projected constraints are shown in Fig.~(\ref{fig:monochromatic_mass_constraints}) for the PTA parameters discussed in Sec.~\ref{subsec:signalSNR}, where the new results, labeled by `DopStoch' and `ShapStoch', are from the stochastic signal caused by the Doppler and Shapiro effects respectively. Shown in orange and green are the reach curves derived in Ref.~\cite{Dror2019a} from deterministic (`Det') or static (`Static') events, corrected to include the subtraction effects. Fig.~(\ref{fig:monochromatic_mass_constraints_opt}) shows the same results as Fig.~(\ref{fig:monochromatic_mass_constraints}) but for more futuristic PTA parameters described in the caption.

We first note that the deterministic signal constraints shown in Fig.~(\ref{fig:monochromatic_mass_constraints}) differ from those in Ref.~\cite{Dror2019a} even for the same sets of PTA parameters.  This is because subtraction of fitted pulsar parameters was neglected previously. To better understand the effect of the subtraction, we look at both the signal and noise power in frequency space. 

\begin{figure}[ht!]
\includegraphics[width=0.8\textwidth]{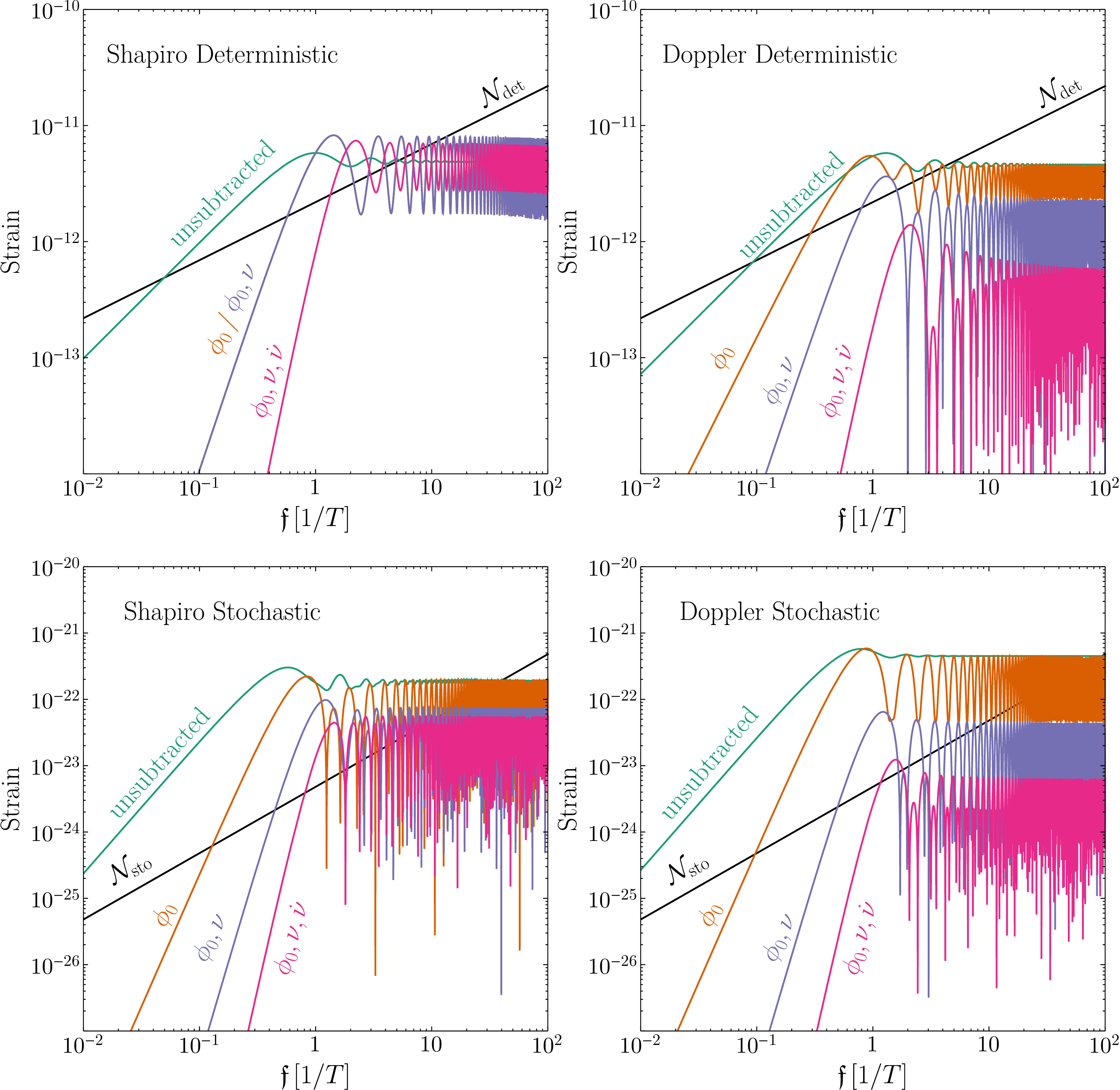}
\caption{Signal strain for the deterministic and stochastic signals from monochromatic PBHs compared to timing noise, $\mathcal{N}$, with parameters $N_P = 200$, $t_\text{rms} = 50 \text{ ns},~\Delta t = 2 \text{ weeks}$ and $T=20~\textrm{years}$. We illustrate the effect of subtraction due to fitting the terms in the timing model, $\phi_0, \nu, \dot{\nu}$. $M = 10^{-3} M_\odot$, $b = 10^{-4} \, \bar{v}_\perp T$ are assumed for the Shapiro signals, $M = 10^{-9} M_\odot$ for the stochastic Doppler signal, and $M = 10^{-8} M_\odot$ for the deterministic Doppler signal.}
\label{fig:power_plot}
\end{figure}

In Fig.~(\ref{fig:power_plot}) we plot the signal and noise strain ($\mathcal{S}$ and $\mathcal{N}$ respectively) for the deterministic and stochastic signals, following the conventions in Ref.~\cite{Moore2015a}, and show the effects of subtracting different terms in the timing model, as in Eq.~\eqref{eq:timingmodel}. We define the signal strains for the deterministic and stochastic signals, $\mathcal{S}_\text{det}, \mathcal{S}_\text{stoch}$ such that
\begin{align}
\text{SNR}_\text{det}^2 & = \, \int d\log{\mathfrak{f}} \frac{\mathcal{S}_{\text{det}}^2(\mathfrak{f})}{\mathcal{N}^2_{\text{det}}(\mathfrak{f})} \\
\text{SNR}_\text{stoch}^2 & = \, \int d\log{\mathfrak{f}}~d\log{\mathfrak{f}'} \frac{\mathcal{S}_{\text{stoch}}^2(\mathfrak{f},\mathfrak{f}')}{\mathcal{N}_{\text{stoch}}(\mathfrak{f}) \mathcal{N}_{\text{stoch}}(\mathfrak{f}')} \, \label{eq:snr_stoch} ,
\end{align}
where the overall $N_P$ dependence for pulsar and Earth terms, is absorbed in $\mathcal{S}_\text{det}, \mathcal{S}_\text{stoch}$, and $\mathfrak{f}$ is used for frequency, to avoid confusion with the dark matter mass fraction, $f$. These definitions allow one to estimate the contribution from different decades in frequency to the SNR using Fig.~(\ref{fig:power_plot}). For comparison with white noise we show the white noise strain, $\mathcal{N}_{\text{det}}(\mathfrak{f}) = \sqrt{\mathfrak{f} \,  \nu^2 t_\text{rms}^2 \Delta t}$, and $\mathcal{N}_{\text{stoch}}(\mathfrak{f}) = \mathfrak{f} \, \nu^2 t_\text{rms}^2 \Delta t$ for an SKA-like PTA.

The deterministic signal strain, shown in the top row of Fig.~(\ref{fig:power_plot}), is from the pulsar term SNR and therefore 
\begin{align}
\mathcal{S}_{\text{det}} = \mathfrak{f} \left| \int dt \, e^{2 \pi i \mathfrak{f} t}h(t) \right| \, ,
\end{align}
where $h$ is the subtracted signal in Eq.~\eqref{eq:subtracted_single_signal_expanded}. The stochastic signal SNR depends on a two dimensional integral, as seen in Eq.~\eqref{eq:snr_stoch}, and therefore the signal strain cannot be plotted as simply as the deterministic signal strain. Instead we show a one-dimensional slice $\mathfrak{f} = \mathfrak{f}'$, where the signal strains for the Doppler (Shapiro) delays, $\mathcal{S}^{\text{stoch}}_{D(S)}$, can be written in terms of a power, $P(\mathfrak{f})$,
\begin{equation}
P(\mathfrak{f}) \equiv \int dt \, dt' \,  e^{2\pi i \mathfrak{f}(t+t')}  R_\text{sub}(t,t') \, .
\label{eq:power_slice}
\end{equation}
The signal strains are then
\begin{align}
\mathcal{S}^{\text{stoch}}_{D(S)}(\mathfrak{f}, \mathfrak{f}) = \sqrt{\frac{N_P}{2}} \mathfrak{f}^2 \left| P_{D(S)}(\mathfrak{f}) \right| \, .
\end{align}
In each panel of Fig.~(\ref{fig:power_plot}) the strain corresponding to the unsubtracted signal is contrasted with the strain from signals with increasingly higher order subtractions corresponding to $\phi^0, \nu,$ and $\dot{\nu}$. Subtraction has the largest effect at frequencies $\lesssim 1/T$, and increasing the number of terms subtracted increases the power law scaling at low frequencies. However, there is also substantial reduction in the strain at large frequencies, although the noise strain is larger and therefore these decades in frequency contribute less to the SNR than the frequencies $\sim 1/T$. For a rough estimate of the SNR one simply needs to estimate the area between the signal and noise strain curves in Fig.~(\ref{fig:power_plot}) near $\mathfrak{f} \sim 1/T$.

We now discuss in more detail how the subtracted signals shown in Fig.~(\ref{fig:power_plot}) were obtained.  Consider first the deterministic Doppler and Shapiro signals. The raw signal, $\delta \phi$, is found by integrating Eqs.~\eqref{eq:shapiro_signal_shape},~\eqref{eq:dop_signal_simple}, and the subtracted signal, $h$, is subsequently computed by Eq.~\eqref{eq:single_subtracted_signal}. The pulsar term SNR from Eq.~\eqref{eq:snr_det_pulsar} for a subtracted and centered ($t_0 = T / 2$) Doppler signal ($\text{SNR}_{\text{det, D, P}}$), in the $b \ll v T$ limit, is given by\footnote{In the $b \ll v T$ limit the Doppler signal is a step function, as seen in Eq.~\eqref{eq:dop_signal_simple}, similar to the gravitational wave memory effect discussed in Ref.~\cite{VanHaasteren2018a}. We also observe that the deterministic SNR is peaked at a signal offset of $t_0/T = 1/2 \pm 1/(2\sqrt{5})$.}
\begin{align}
\text{SNR}_{\text{det}, D, P}\left( t_0 = \frac{T}{2} \right) & = \frac{G M}{16 \sqrt{3}} \frac{T^\frac{3}{2}}{t_\text{rms} \sqrt{\Delta t}} \underset{\{I, i\}}{\max} \left[ \frac{\left| \vec{\hat{d}}_I \cdot \vec{\hat{b}}_i \right|}{b_i v_i} \right] \, ,
\label{eq:corrected_Dop_pulsar}
\end{align}
where the maximum is taken over all events, $i$, in all the pulsars, $I$. This subtracted result is a factor of $8$ smaller compared to Ref.~\cite{Dror2019a} where no subtraction was done. Likewise, the subtracted pulsar term SNR for a subtracted  and centered ($t_0 = T/2$) Shapiro signal ($\text{SNR}_{\text{det}, S, P}$), in the $b \ll \overline{v}_\perp T$ limit, is given by
\begin{align}
\text{SNR}_{\text{det}, S, P}\left( t_0 = \frac{T}{2} \right) \approx 1.33 \times \frac{G M}{t_\text{rms}} \sqrt{\frac{T}{\Delta t}} \, ,
\end{align}
which is a factor of $0.24$ smaller in Ref.~\cite{Dror2019a}. Subtraction has a larger effect on the Doppler delay because the Shapiro signal is much more peaked, and therefore less susceptible to subtraction. The projected constraints for the deterministic signals shown in our reach plots have been appropriately rescaled to account for this $\mathcal{O}(1)$ change in the SNR. Also note that we have considered a centered signal, $t_0 = T/2$, to mirror the analysis done in \cite{Dror2019a}. A more accurate analysis would include the $t_0$ dependence when computing the SNR via a Monte Carlo (MC) simulation; we expect this effect to be small, however, as the SNR only decreases rapidly when $t_0$ is near the observation edge: $t_0 = 0$ or $t_0 = T$.

Lastly, in Ref.~\cite{Dror2019a} the Doppler Earth term was considered subdominant compared to the pulsar term and ignored. This is true for PBHs, as the PBH closest to the Earth is farther than the one closest to any pulsar. However, this effect is compensated by a factor of $N_P$ in the Earth term SNR, and for more diffuse subhalos this increase in the minimum impact parameter extends the reach, as more subhalo mass is contained within the impact parameter. The Earth term will therefore be more sensitive to diffuse subhalos relative to the pulsar term, as seen in Fig.~(\ref{fig:monochromatic_mass_constraints}). The Earth term SNR, for a centered Doppler signal, is given by
\begin{align}
\text{SNR}_{\text{det, D, E}}\left( t_0 = \frac{T}{2} \right) & = \frac{\sqrt{N_P}}{48} \frac{G M T^\frac{3}{2}}{t_\text{rms} \sqrt{\Delta t}} \underset{\{ i \}}{\max} \left[ \frac{1}{b_i v_i} \right] \, ,
\end{align}
where $i$ is over all of the events near the Earth, and the average over the pulsar positions gives a factor of $1/\sqrt{3}$ relative to Eq.~\eqref{eq:corrected_Dop_pulsar}. 

Next we discuss and derive in detail the features from the new analysis of the stochastic signal (curves labeled `stoch' in Fig.~(\ref{fig:monochromatic_mass_constraints})), which can be understood from the distance scales discussed in the previous subsection.
First, the right-hand side of the `stoch' reach curves in Fig.~(\ref{fig:monochromatic_mass_constraints}), $f_\dm^R$, is derived from the requirement that events transit the Earth-pulsar system during the observation time, {\em i.e.} $b_{\rm min}^{D(S)} < \overline{v}_{(\perp)} T$. Utilizing Eq.~\eqref{eq:b_min_statistics} we obtain:
\begin{align}
f^R_S & \sim 0.5 \, \left( \frac{M}{10^{-3} M_\odot} \right) \left( \frac{20 \text{ year}}{T} \right)^2 \left( \frac{5 \kpc}{z_0} \right)\\
f^R_D & \sim 0.2 \, \left( \frac{M}{10^{-9} \, M_\odot} \right) \left( \frac{20 \text{ year}}{T} \right)^3\, .
\label{eq:bestmass}
\end{align}

The opposite (left-hand) side of the `Stoch' constraints are derived by the strength of the SNR. Evaluating Eqs.~\eqref{eq:R_sh_final},~\eqref{eq:R_dop_final},
\begin{align}
R_S(t, t') & = 4 \pi \, f \, G^2 \rho_\text{DM} M \nu^2 \overline{v}_\perp^2 z_0 T^2 \int_0^\frac{t}{T} \int_0^\frac{t'}{T} dx_1 dx_2 \, \text{log}\left( \frac{4 + (x_1 - x_2)^2}{4 \left( \frac{b_{S,\text{min}}}{\overline{v}_\perp T} \right)^2 + (x_1 - x_2)^2} \right) \nonumber \\
& \equiv  4 \pi \, f \, G^2 \rho_\text{DM} M \nu^2 \overline{v}_\perp^2 z_0 T^2 \, C\left(\frac{t}{T}, \frac{t'}{T}, \frac{b^S_\text{min}}{\overline{v}_\perp T} \right) \label{eq:R_S_mono_PBH}\\
R_D^{\, IJ}(t, t') & = \frac{4 \pi \, f \, G^2 \rho_\text{DM} M \nu^2}{\overline{v}} \left( \vec{\hat{d}}_I \cdot \vec{\hat{d}}_J \right) B(t, t') \text{log}\left( \frac{v T}{b_\text{min}^D} \right) \, ,\label{eq:R_D_mono_PBH}
\end{align}
which must then be substituted in to Eq.~\eqref{eq:sub_auto_corr_main} before computing the SNR with Eqs.~\eqref{eq:SNR_P_simple},~\eqref{eq:SNR_E_simple}. Beginning with the Shapiro stochastic signal, which has an unsubtracted correlator given in Eq.~\eqref{eq:R_S_mono_PBH}, the SNR in Eq.~\eqref{eq:SNR_P_simple} is approximately, 
\begin{align}
\text{SNR}_S& \approx \left( 9.4 \times 10^{-2} \right) \, f \, \frac{\sqrt{N_P} G^2 \rho_\dm M \overline{v}_\perp^2 z_0 T^3}{t_\text{rms}^2 \Delta t} \, ,
\end{align}
in the $b_\text{min} \ll \overline{v}_\perp T$ limit. Setting $\text{SNR}_S = 2$, gives the left-hand side of the `ShapStoch' constraint, $f_{\dm(S)}^L$,
\begin{align}
f_{S}^L \approx 2.3 \, \left( \frac{200}{N_P} \right)^\frac{1}{2} \left( \frac{10^{-4} M_\odot}{M} \right) \left( \frac{t_\text{rms}}{50 \text{ ns}} \right)^2 \left( \frac{5 \kpc}{z_0} \right) \left( \frac{20 \text{ year}}{T} \right)^3 \, .
\end{align}
The left-hand side of the `DopStoch' curves in Fig.~(\ref{fig:monochromatic_mass_constraints}) are derived similarly. The average over the pulsar positions contributes a factor of
\begin{align}
\left\langle \left( \vec{\hat{d}}_I \cdot \vec{\hat{d}}_J\right)^2 \right\rangle_\mathcal{P} = \frac{1}{3} \, ,
\end{align}
and the SNR in Eq.~\eqref{eq:SNR_E_simple} is then,
\begin{align}
\text{SNR}_D 
& \approx \left( 1.4 \times 10^{-3} \right) f \, \rho_\dm \frac{N_P G^2 T^4 M}{\Delta t \, t_\text{rms}^2 \overline{v}} \mylog{\frac{\overline{v} T}{b^D_\min}} \, . \label{eq:SNR_mono_dop}
\end{align}
The constraint is again derived from $\text{SNR}_D = 2$. There is no simple scaling law as there is $f$ dependence inside the logarithm. 

In deriving analytic results for the stochastic signals, we are using the expressions for $b_\min$ from Eq.~\eqref{eq:b_min_statistics}, and we justify their use here. The SNRs in Eqs.~\eqref{eq:SNR_P_simple},~\eqref{eq:SNR_E_simple} have been calculated by averaging over the subhalo random variables, \textit{e.g.} $\vec{b}_i$. The subtlety is that the average can be skewed by unlikely values. For example, the Doppler delay SNR$^2 \propto  \langle b^{-2} \rangle$, diverging with the lower cut-off on the $b$ integral; even though small values of $b$ are unlikely, their effect on the SNR is large enough to skew the average.  This calls into question the robustness of our analytic prescription with $b_{\rm min}$ calculated from Eq.~\eqref{eq:b_min_statistics}. 

\begin{figure}[t]
\includegraphics[width=0.8\textwidth]{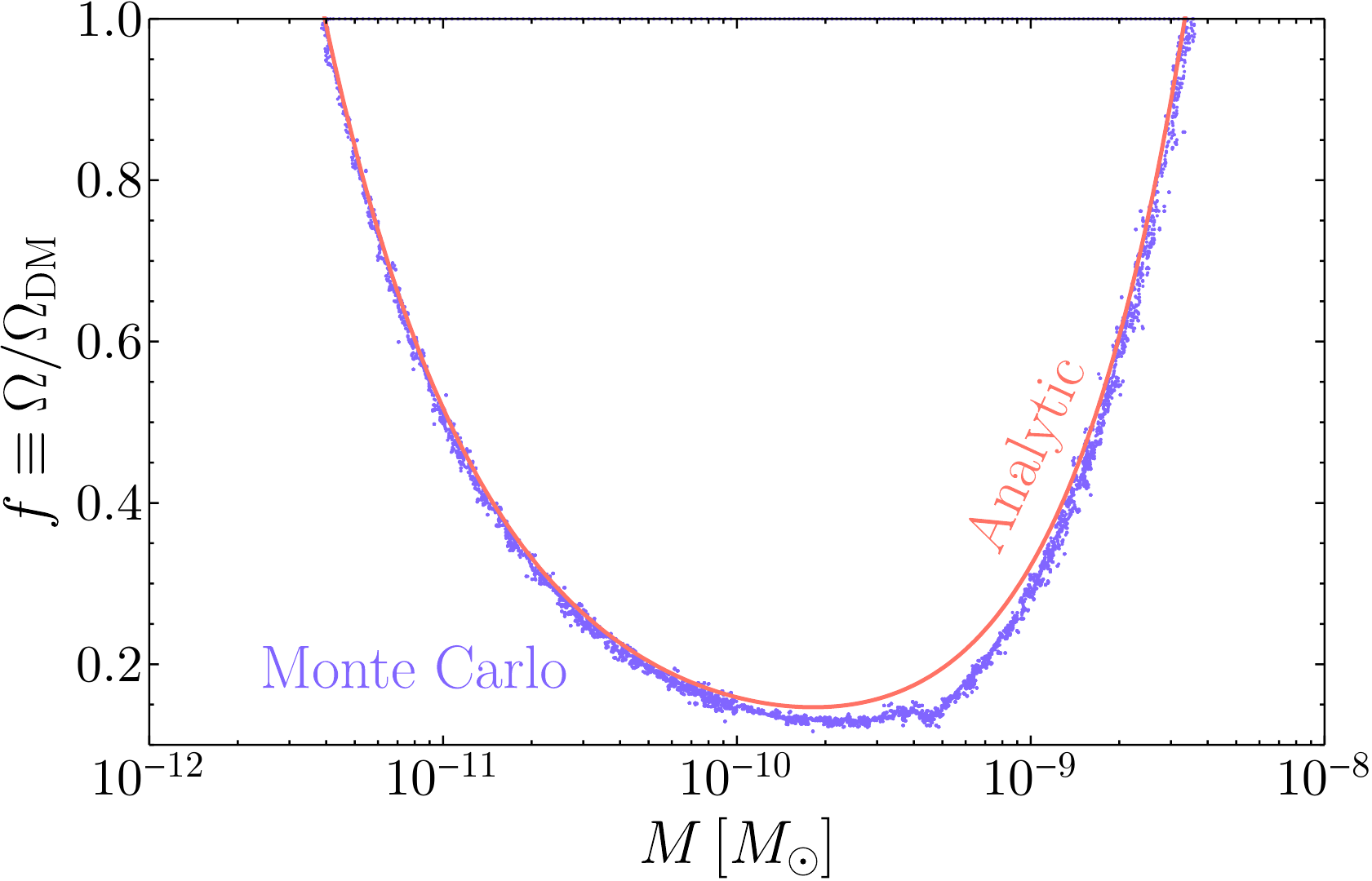}
\caption{Comparison of constraints for the Doppler stochastic signal using the Monte Carlo (MC) and analytic approaches in the PBH limit. The MC derives constraints from the simulated $10$\th percentile SNR, whereas the analytic constraint imposes $b > b_\text{min}^D$, where $b_\text{min}^D$ is the $90$\th percentile minimum impact parameter. Constraints are created assuming $v = 10^{-3}$ and PTA parameters of $N_P = 200, t_\text{rms} = 10 \text{ ns}, \Delta t = 1 \text{ week}$.}
\label{fig:monochromatic_MCvAnalytic}
\end{figure}

The solution is to calculate a skew-independent statistic of the signal, such as a percentile, which we obtain via a Monte Carlo (MC) simulation. We find good agreement between the reach calculated analytically, with $b_\min$ from Eq.~\eqref{eq:b_min_statistics}, and the $10$\th percentile SNR computed from the MC. The results are compared in Fig.~(\ref{fig:monochromatic_MCvAnalytic}). Because of the strong agreement, we will proceed to use the analytic results in the remainder of this paper.

\subsubsection{NFW Subhalo}
\label{sec:mono_mass_dist_co}

We now turn to less concentrated subhalos and, for concreteness, consider an NFW density profile,
\begin{align}
\rho(r, c)  = \frac{4\rho_s(c)}{(r/r_s)(1 + r/r_s)^2}, 
\label{eq:nfwprofile}
\end{align}
where $r_s$ is the scale radius, and the scale density $\rho_s(c) = \rho(r_s,c)$ is given by,
\begin{align}
\rho_s(c) & =\frac{50 c^3 \rho_c}{3 (\log (c+1)-c/(1+c))},
\label{eq:densc}
\end{align}
where $\rho_c$ is the critical density and $c$ is the concentration parameter. As emphasized in Eq.~\eqref{eq:rs}, for a given subhalo mass $M$, larger concentration parameters lead to more compact subhalos, with the PBH limit $c \rightarrow \infty$. $N$-body simulations of CDM subhalos indicate that $10 \lesssim c \lesssim 100$, but more concentrated subhalos can be formed from earlier collapse, as $c \propto 1 + z_\text{col}$, where $z_\text{col}$ is the collapse redshift.  

With the density profile defined, the steps to calculating the constraints are the same as that of a point-like subhalo but now the form factor ${\cal F}$ appears in Eqs.~\eqref{eq:R_sh_final},~\eqref{eq:R_dop_final};  the results are similar to those in Eqs.~\eqref{eq:R_S_mono_PBH},~\eqref{eq:R_D_mono_PBH}, except now we do not take $\mathcal{F} \rightarrow 1$: 
\begin{align}
R_S(t, t') & = 32 \pi G^2 M \,f \, \rho_\dm \nu^2 z_0 \int db \, b \, A\left( t, t', \frac{b}{\overline{v}_\perp} \right) \left( \mathcal{F}\left( \frac{b}{r_v}, c \right) \right)^2 \label{eq:R_S_mono_CO}\\
R_D^{\, IJ}(t, t') & = \frac{4\pi G^2 M\,f \, \rho_\dm \nu^2}{\overline{v}} \left( \vec{\hat{d}}^I \cdot \vec{\hat{d}}^J \right) B(t, t') \int \frac{db}{b}\, \left( \mathcal{F}\left(\frac{b}{r_v}, c\right) \right)^2 \label{eq:R_D_mono_CO} \, .
\end{align}

\begin{figure}[t]
\includegraphics[width=0.6\textwidth]{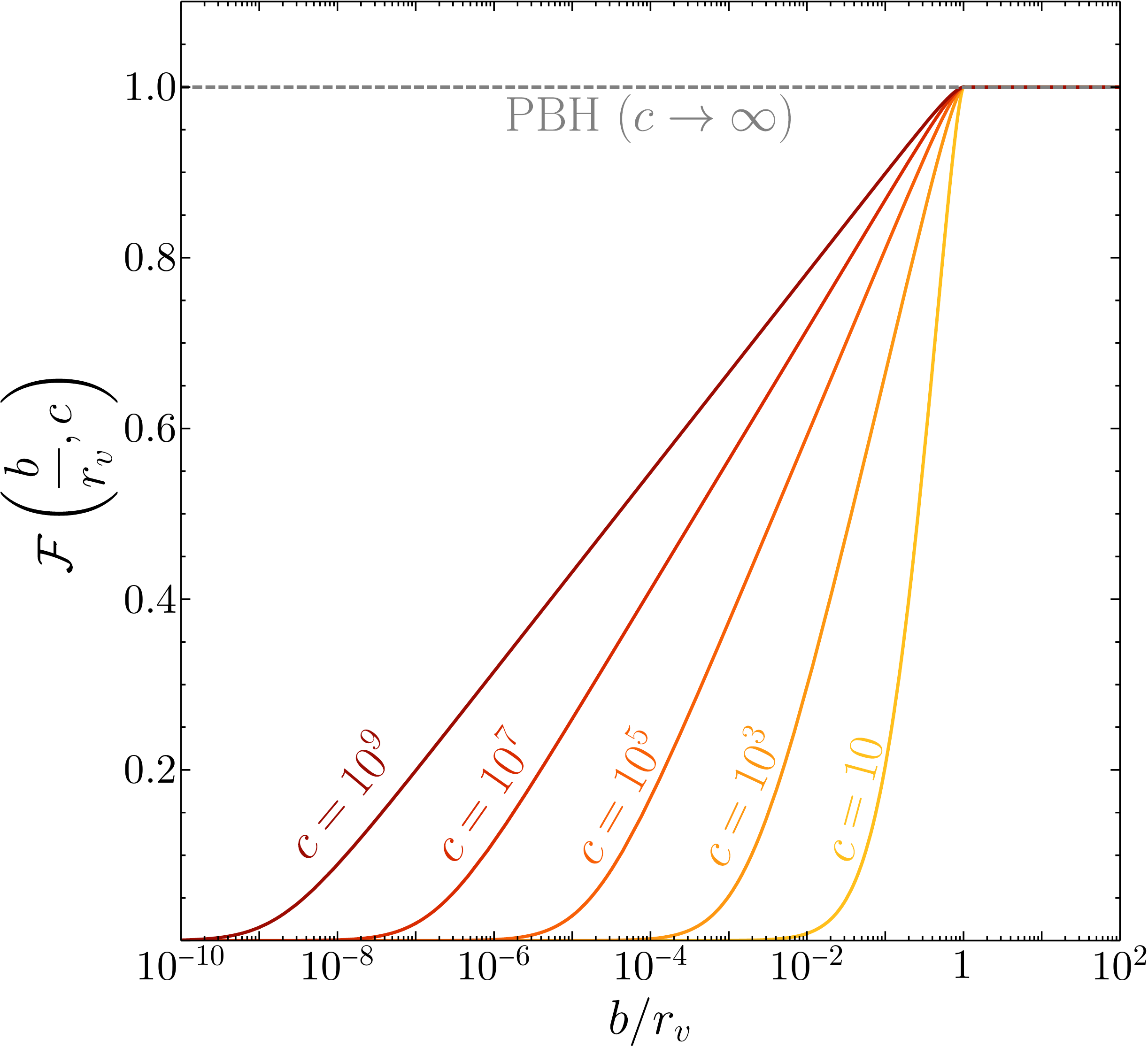}
\caption{Subhalo form factor ${\cal F}$ for various concentration parameters $c$, as a function of the subhalo impact parameter $b$ and virial radius $r_v$. When $b/r_v > 1$, the subhalo is point-like and ${\cal F} \rightarrow 1$.  As $b/r_v$ drops below 1, ${\cal F}$ drops only logarithmically with $b/r_v$ until $b/r_v < 1/c$, where the form factor rapidly goes to zero. A PTA can only observe subhalos with $b< b_{\rm max} = v T$, set by the PTA observing time $T$. Therefore more massive subhalos, having larger virial radii, can only be constrained if they have sufficiently large concentration parameter. 
}
\label{fig:formfactor}
\end{figure} 

The result of computing the constraints on the dark matter mass fraction, $f$, are shown in Fig.~(\ref{fig:monochromatic_mass_constraints}) for $c=10,~100,~10^4$. Because the signal depends on the integral from $b_{\rm min}$ to $b_{\rm max}$, the difference in constraints between finite $c$ and $c \rightarrow \infty$ can be understood from the behavior of the form factor $\mathcal{F}$ over this range of impact parameters. This is shown in Fig.~(\ref{fig:formfactor}) (obtained from Eq.~\eqref{eq:F_int} and simplified analytically in Appendix~\ref{app:form_factor}) as a function of $b/r_v$ and $c$.  ${\cal F} \rightarrow 1$ when $b/r_v > 1$ and the masses can be treated as point-like.  As long as $b/r_v > 1/c$, ${\cal F}$ remains relatively large. Only once $b/r_v < 1/c$ does ${\cal F}$ drop rapidly and the signal becomes very weak. This relative insensitivity to the subhalo radius allows PTAs to constrain a wide range of concentration parameters.

For example, the `DopStoch' constraints are relatively $c$-independent because, even for $c \gtrsim 10$, $r_v/c \lesssim b_\text{min}^D$ (see Fig.~(\ref{fig:relevant_distances})), such that $\mathcal{F} \sim 1$ over the integration region. On the other hand, the Shapiro search is only sensitive to larger mass subhalos (as seen in Fig.~(\ref{fig:monochromatic_mass_constraints})) that have larger radii. $b/r_v$ is typically thus much smaller, and ${\cal F}$ is rapidly suppressed, as shown in Fig.~(\ref{fig:formfactor}), so that the Shapiro search has much less reach to low concentration subhalos.   

In summary, in order for subhalos to be sufficiently compact to be observable by PTAs, we require their scale radius be smaller than the radius of the observing volume, $ r_s < b_\text{max} $, which is only satisfied for large concentration parameters: 
\begin{align}
c \gg 4 \times 10^{3} \left( \frac{20 \text{ yr}}{T} \right) \left( \frac{M}{M_\odot} \right)^\frac{1}{3}.
\end{align}
Overall, this means that PTAs are particularly powerful probes for low concentration subhalos with $M \lesssim 10^{-6}~M_\odot$.

\begin{figure}[ht]
\includegraphics[width=\textwidth]{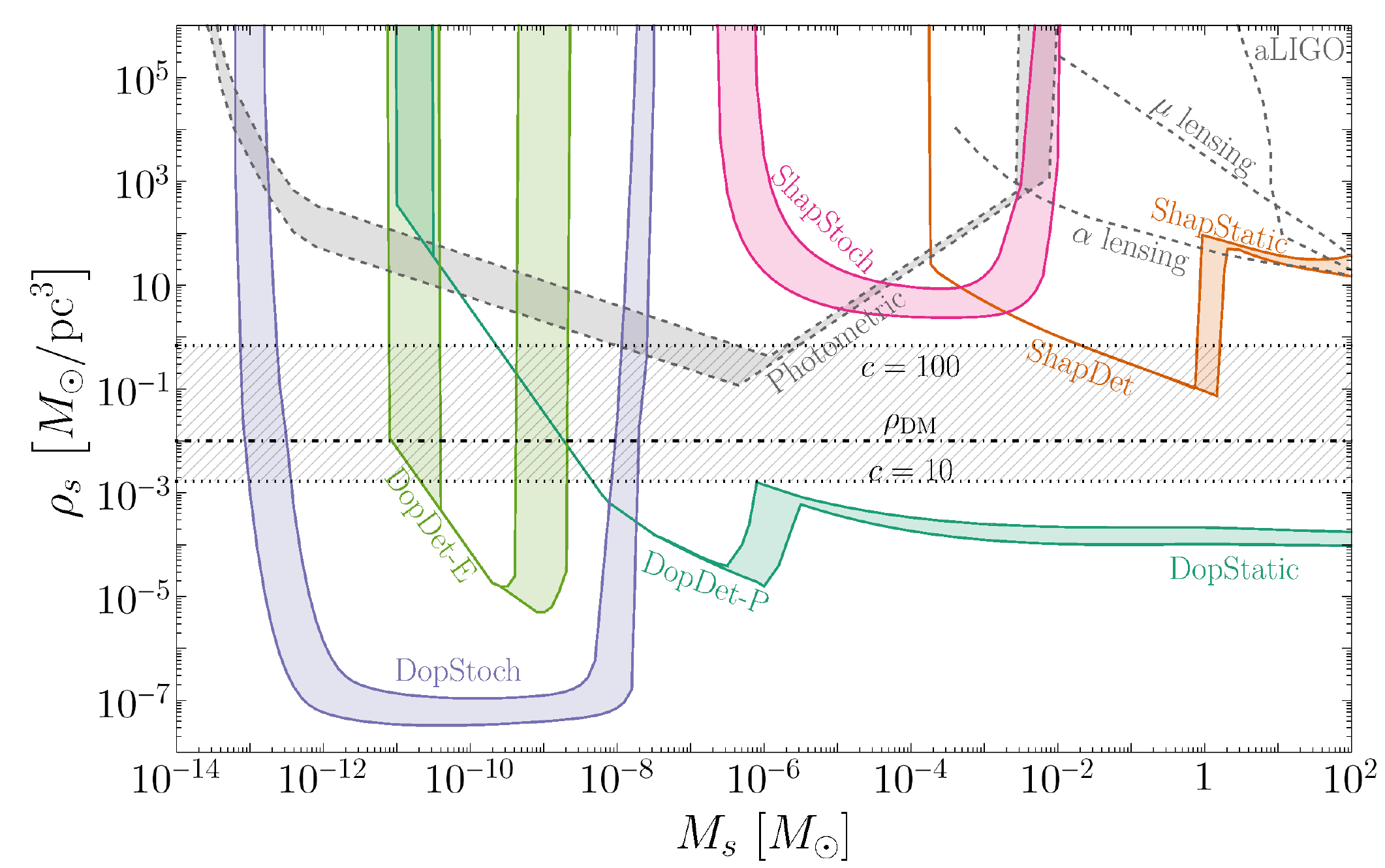}
\caption{PTA constraints on a monochromatic mass distribution of `core-only' subhalos with scale density $\rho_s$, as a function of core mass $M_s$ arising from optimistic PTA parameters. PTA constraints have the same color scheme as in Fig.~(\ref{fig:monochromatic_mass_constraints_opt}). To show the dependence on the dark matter fraction $f$ in such subhalos, we show a band with $0.3 \leq f \leq 1$. Scale densities corresponding to $10\le c\le1$ in the $\Lambda \textrm{CDM}$ range are shown as a hatched region (though note this is not the $\Lambda$CDM model, which features a broad spectrum of subhalo masses). Halos  below the dot-dashed-black line corresponding to $\rho_s=\rho_{\rm DM}$ cannot make up $f=1$ . Also shown in dashed-gray are projections from photometric lensing \cite{Dai:2019lud,Blinov:2019jqc}, $\alpha$ and $\mu$ lensing from astrometric lensing in Gaia data \cite{VanTilburg:2018ykj, Mondino:2020rkn, Mishra-Sharma:2020ynk} and diffraction of BH mergers observable in aLIGO\cite{Dai:2018enj,Arvanitaki:2019rax}.}
\label{fig:densityplot}
\end{figure}

In Refs.~\cite{Arvanitaki:2019rax,Blinov:2019jqc} the effect of tidal stripping on diffuse subhalos was incorporated by assuming that only cores survive until late times. This was modeled by an abrupt fall-off in density outside the scale radius $r_s$ in Eq.~\eqref{eq:nfwprofile}. For these subhalos, $M = M_s$ where $M_s$ is the mass contained inside the radius $r_s$. Constraints from different probes were projected in the $\rho_s$ versus $M_s$ plane, where $\rho_s = \rho(r_s)$ is the scale density, for monochromatic subhalos which make up a fraction $f$ of the dark matter. For a direct comparison with other futuristic proposals, we show constraints from PTAs with optimistic pulsar parameters in Fig.~(\ref{fig:densityplot}). For each probe, we show two contours corresponding to the minimum scale density that can be probed for a particular core mass, $M_s$, for $f=1$ and $f=0.3$. Also shown are projections from photometric lensing~\cite{Dai:2019lud}, as computed in \cite{Blinov:2019jqc}, $f=1$ constraints from astrometric $\alpha$ and $\mu$ lensing from Gaia data \cite{VanTilburg:2018ykj, Mondino:2020rkn, Mishra-Sharma:2020ynk}, and diffraction of gravitational wave from BH mergers observable at aLIGO~\cite{Dai:2018enj,Arvanitaki:2019rax} in dashed gray. The hatched region corresponds to subhalos with $10\le c \le 100$, to account for $\Lambda$CDM-like subhalos at masses much lower than those typically found in simulations~\cite{Moline:2016pbm,Wang2019a}. The black dot-dashed line corresponds to the local DM density; subhalos cannot make up all of dark matter, without sufficiently overlapping, below this line. The robustness of constraints with respect to the concentration parameter, as seen in Figs.~(\ref{fig:monochromatic_mass_constraints}),~(\ref{fig:monochromatic_mass_constraints_opt}), translates to sensitivities to very small $\rho_s$ in Fig.~(\ref{fig:densityplot}). Remarkably, with optimistic PTA parameters, a combination of the different Doppler constraints will be sensitive to a monochromatic mass distribution with even the most diffuse cores, in a mass window spanning as low as $10^{-13} M_\odot$ to well above a solar mass. 

\subsection{Extended Halo Mass Functions}
\label{sec:ext_mass_dist}

We now consider a mass distribution that is not simply monochromatic, focusing on the highly relevant case of the Cold Dark Matter (CDM) paradigm where scale invariant perturbations are seeded by inflation. The (nearly) scale invariant nature of the perturbations fairly firmly fixes the spectrum of the Halo Mass Function (HMF), which can be parameterized as 
\begin{align}
F(M) \equiv \frac{dn}{d \log{M}} & = \frac{f \rho_\dm}{\mathcal{N}(M_\min, M_\max, 1 - \beta)} M^{1 - \beta} \Theta(M_\max - M) \Theta(M - M_\min) \label{eq:CDMHMF}\\
\mathcal{N}(M_\min, M_\max, \alpha) & \equiv \int_{M_\min}^{M_\max} M^\alpha dM = 
	\begin{cases}
		\mylog{\frac{M_\max}{M_\min}} & \alpha = -1 \\
		\frac{1}{\alpha + 1} \left( M_\max^{1 + \alpha} - M_\min^{1 + \alpha} \right) & \alpha \neq -1
	\end{cases},
	\label{eq:CDMHMF_norm}
\end{align}
where the overall normalization, $\mathcal{N}$, is found by requiring that the expected matter density is equal to its measured value, $f \rho_\dm = \int dM \, F(M)$. The total number of subhalos within a volume $V$ is given by,
\begin{align}
N = V \int_{M_\min}^{M_\max} \frac{dn}{dM} dM = f \rho_\dm V \frac{\mathcal{N}(M_\min, M_\max, -\beta)}{\mathcal{N}(M_\min, M_\max, -\beta + 1)}.
\end{align}

One can estimate $\beta$ with the standard Press-Schechter theory~\cite{Press1974a}, assuming a scale invariant primordial power spectrum. For large $k$ the power spectrum today scales as $\sim k^{-3}$, up to the free-streaming scale (which sets $M_\min$), corresponding to $\beta = 2$. This scale invariant spectrum gives equal mass density in equal logarithmic intervals. 

Cosmological $\Lambda$-CDM $N$-body simulations have, however, made more precise estimates which indicate that $\beta \approx 1.9$~\cite{Springel2008a,Fiacconi2016a}. As innocent as this difference seems, it has a large impact on the constraints that can be placed at low subhalo masses where PTA constraints are most powerful. This is because, for $\beta = 1.9$, the mass density is dominated by the large mass subhalos. Since $M_\min, M_\max$ and $\beta$ are largely model-dependent, we allow them to vary, but choose parameters that do not dramatically vary from a scale invariant spectrum.   

\begin{figure}[t]
\includegraphics[width=\textwidth]{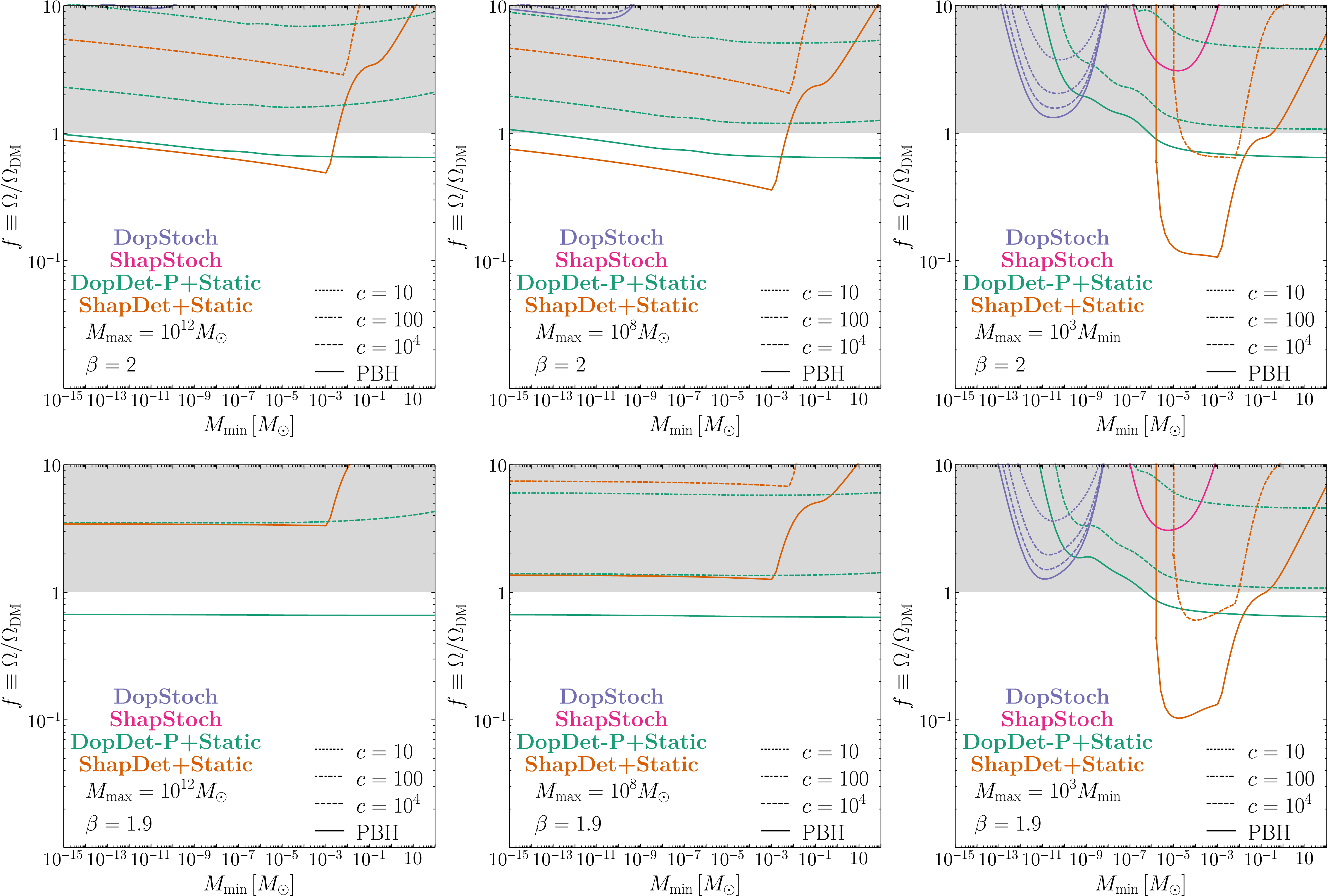}
\caption{Constraint on the fraction of dark matter $f = \Omega / \Omega_\dm$ in a Halo Mass Function (HMF) parameterized by $M_\min, M_\max$ and $\beta$, as in Eq.~\eqref{eq:CDMHMF_norm}. We assume SKA-like PTA parameters of $N_P = 200$, $T = 20 \text{ yr}$, $t_\text{rms} = 50 \text{ ns}$, $\Delta t = 2 \text{ weeks}$, $z_0 = 5 \text{ kpc}$. We show constraints from four different signal regimes: combined pulsar term deterministic and static Doppler (DopDet-P+Static), combined deterministic and static Shapiro (ShapDet+Static), stochastic Doppler (DopStoch), stochastic Shapiro (ShapStoch), with four different concentrations $c = 10, 100, 10^4$, and the PBH-like $c \rightarrow \infty$.}
\label{fig:extendedHMF}
\end{figure}

\begin{figure}[ht!]
\includegraphics[width=\textwidth]{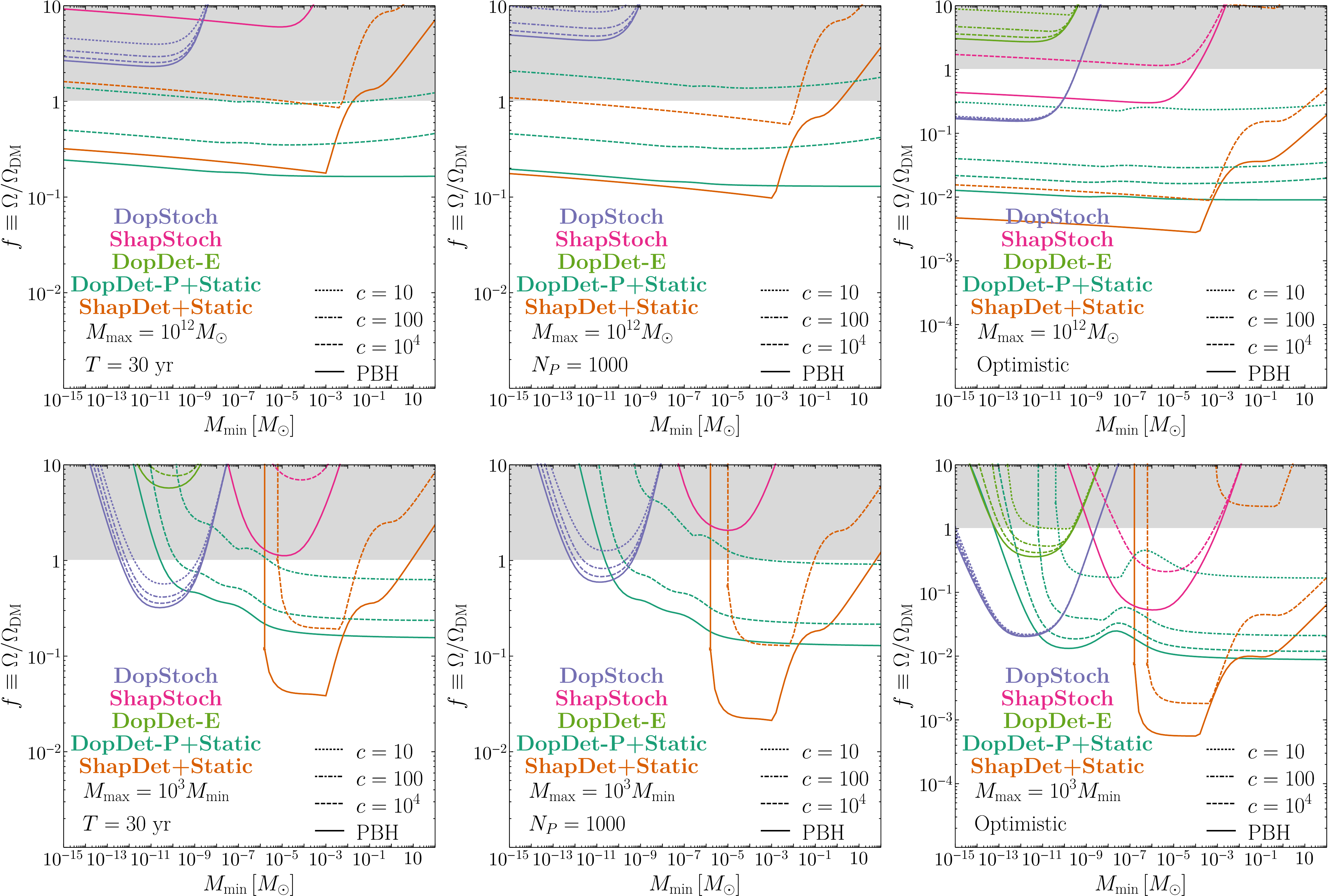}
\caption{Constraint on the fraction of dark matter $f = \Omega / \Omega_\dm$ in a Halo Mass Function (HMF) parameterized by $M_\min, M_\max$ and $\beta = 2$, as in Eq.~\eqref{eq:CDMHMF_norm}.  In the three columns we vary the PTA capability parameters to show what will be necessary to reach a CDM-like HMF.  The left column assumes an SKA-like PTA with $T = 30$ year; the middle column an SKA-like PTA with $N_P = 1000$; the right column shows a futuristic PTA with optimistic parameters: $N_P = 1000$, $T = 30 \text{ yr}$, $t_\text{rms} = 10 \text{ ns}$, $\Delta t = 1 \text{ week}$, $z_0 = 10 \text{ kpc}$. As in Fig.~(\ref{fig:extendedHMF}) we show constraints from four different signal regimes: combined deterministic and static Doppler pulsar term (DopDet-P+Static), combined deterministic and static Shapiro (ShapDet+Static), stochastic Doppler (DopStoch), stochastic Shapiro (ShapStoch). Additionally, the deterministic Earth term Doppler DopDet-E is shown. Four choices of concentration parameter $c = 10, 100, 10^4$, and the PBH limit $c \rightarrow \infty$ are shown.}
\label{fig:extendedHMF_optimistic}
\end{figure}

$N$-body simulations, having dark matter only, favor constituent subhalos with an NFW profile.\footnote{Baryons tend to change these profiles near the core of the subhalo, but the small subhalos that we consider here do not hold baryons} Furthermore, CDM subhalos, obtained from galactic simulations~\cite{Wang2019a,Ludlow2014a,Springel2008a,Diemer2019a}, are typically quoted to have concentration parameters with $c \gtrsim 50$ for subhalos below $M_\odot$.  These models, however, are usually obtained from simulation data with $M \gtrsim M_\odot$. More recent simulations which study lower mass subhalos suggest that, below $M \sim 10^{-3} \, M_\odot$, the concentration parameter decreases with decreasing mass~\cite{Wang2019a}. While the `Doppler-stoch' search will have some reach for $c \sim 10$, it will only be at masses much smaller than the typical minimum mass of CDM subhalos from WIMP dark matter which have $M_\min \sim 10^{-6} \, M_\odot$. Given the uncertainty on the concentration parameters of low mass subhalos, we show our results in Figs.~(\ref{fig:extendedHMF}),~(\ref{fig:extendedHMF_optimistic}) for $c = 10, 100, 10^4$, and the $c \rightarrow \infty$, PBH limit for comparison.

\begin{figure}[t]
\includegraphics[width=\textwidth]{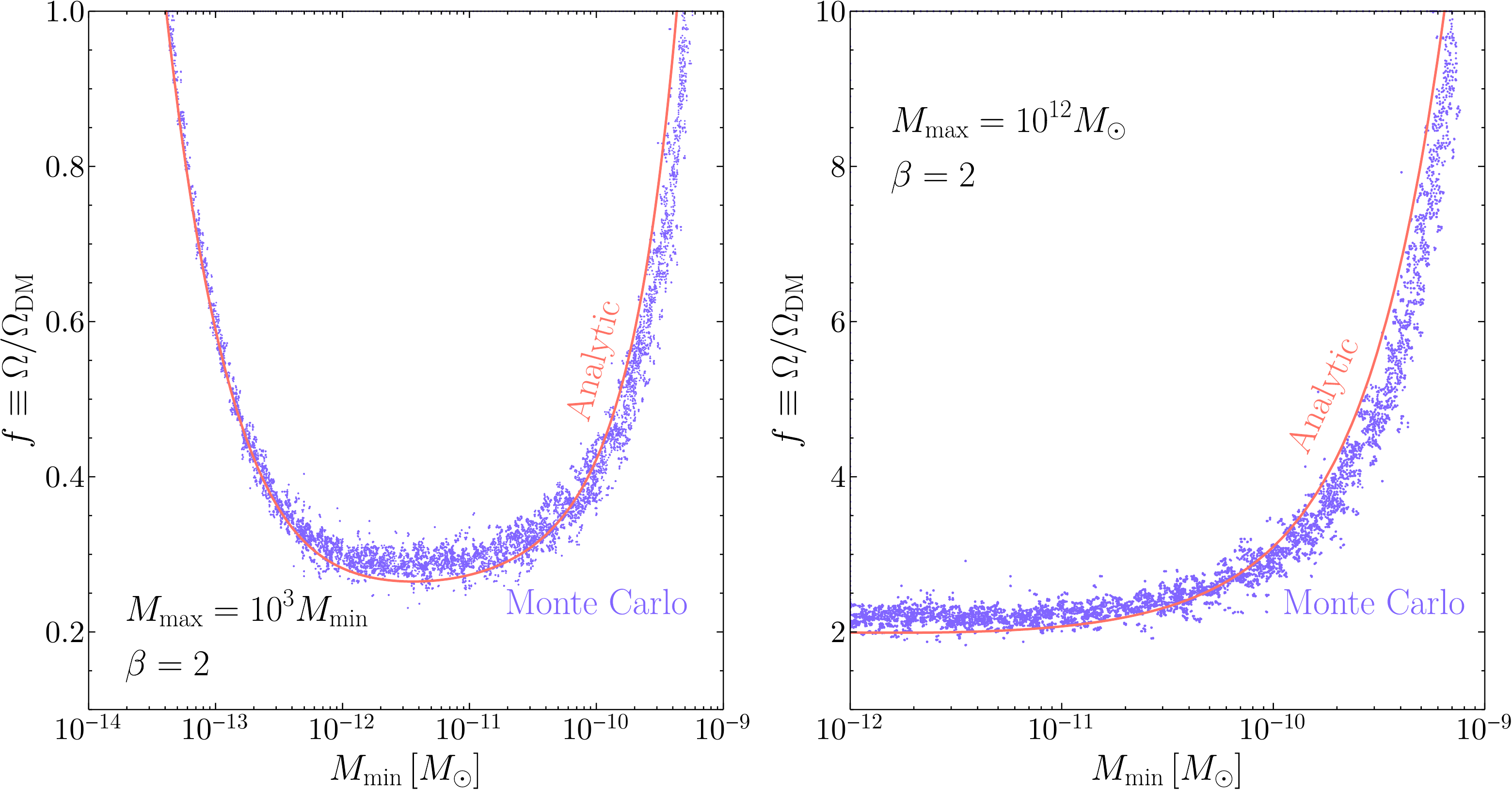}
\caption{Comparison of constraints for the `DopStoch' signal using the Monte Carlo (MC) and analytic approach from Ref.~\cite{Carr2017a}. The MC calculates constraints from the the $10$\th percentile SNR, while the analytic approach re-integrates the monochromatic results using Eq.~\eqref{eq:carr}. The two methods are nevertheless in good agreement. Constraints are again created assuming $v = 10^{-3}$ and PTA parameters of $N_P = 200, t_\text{rms} = 10 \text{ ns}, \Delta t = 1 \text{ week}$.}
\label{fig:extended_MCvAnalytic}
\end{figure}

The constraints shown in Figs.~(\ref{fig:extendedHMF}),~(\ref{fig:extendedHMF_optimistic}) were derived from the monochromatic mass distribution, following Ref.~\cite{Carr2017a}. The method advocated there re-weights the monochromatic distribution constraints shown in Fig.~(\ref{fig:monochromatic_mass_constraints}), $f_\text{mono}(M)$, according to the relation
\begin{equation}
\frac{\rho_{\text{DM}}}{f}=\int dM \frac{F(M)}{f_{\rm mono} (M)} \, .
\label{eq:carr}
\end{equation}
To check the validity of this approximation, we compare the results obtained utilizing this analytic prescription with a Monte Carlo, in Fig.~(\ref{fig:extended_MCvAnalytic}).  We see good agreement for more than one HMF, and proceed to use this analytic formula in our main results, Figs.~(\ref{fig:extendedHMF}),~(\ref{fig:extendedHMF_optimistic}). 

The first row in Fig.~(\ref{fig:extendedHMF}) shows constraints for $\beta = 2$, and the second $\beta = 1.9$. We take $M_\text{max}= 10^{12} M_\odot$ (the Milky Way galaxy mass) and $10^{8} M_\odot$ to show the dependence on $M_\text{max}$ in the left and middle columns, and lastly $M_\max=10^3 M_\text{min}$, in the right column.  Such narrow HMFs can be produced in theories with peaks in the primordial power spectrum \cite{Hogan1988a,Kolb:1993zz,Kolb:1993hw,Zurek2007a,Buschmann:2019icd,Arvanitaki:2019rax,Graham:2015rva,Erickcek:2011us,Barenboim:2013gya,Fan:2014zua}.  We will use a shorthand for the variety of search types: `DopDet-P+Static' for the combined deterministic and static Doppler searches that involve the pulsar term, `ShapDet+Static' for the combined deterministic and static Shapiro searches, `DopStoch' for the stochastic Doppler signal, and `ShapStoch' for stochastic Shapiro signal. The `DopDet-P+Static' and `ShapDet+Static' curves were derived in Ref.~\cite{Dror2019a} and corrected due to the subtraction procedure here, while `DopStoch' and `ShapStoch' were derived here. 

\begin{figure}[t]
\includegraphics[width=\textwidth]{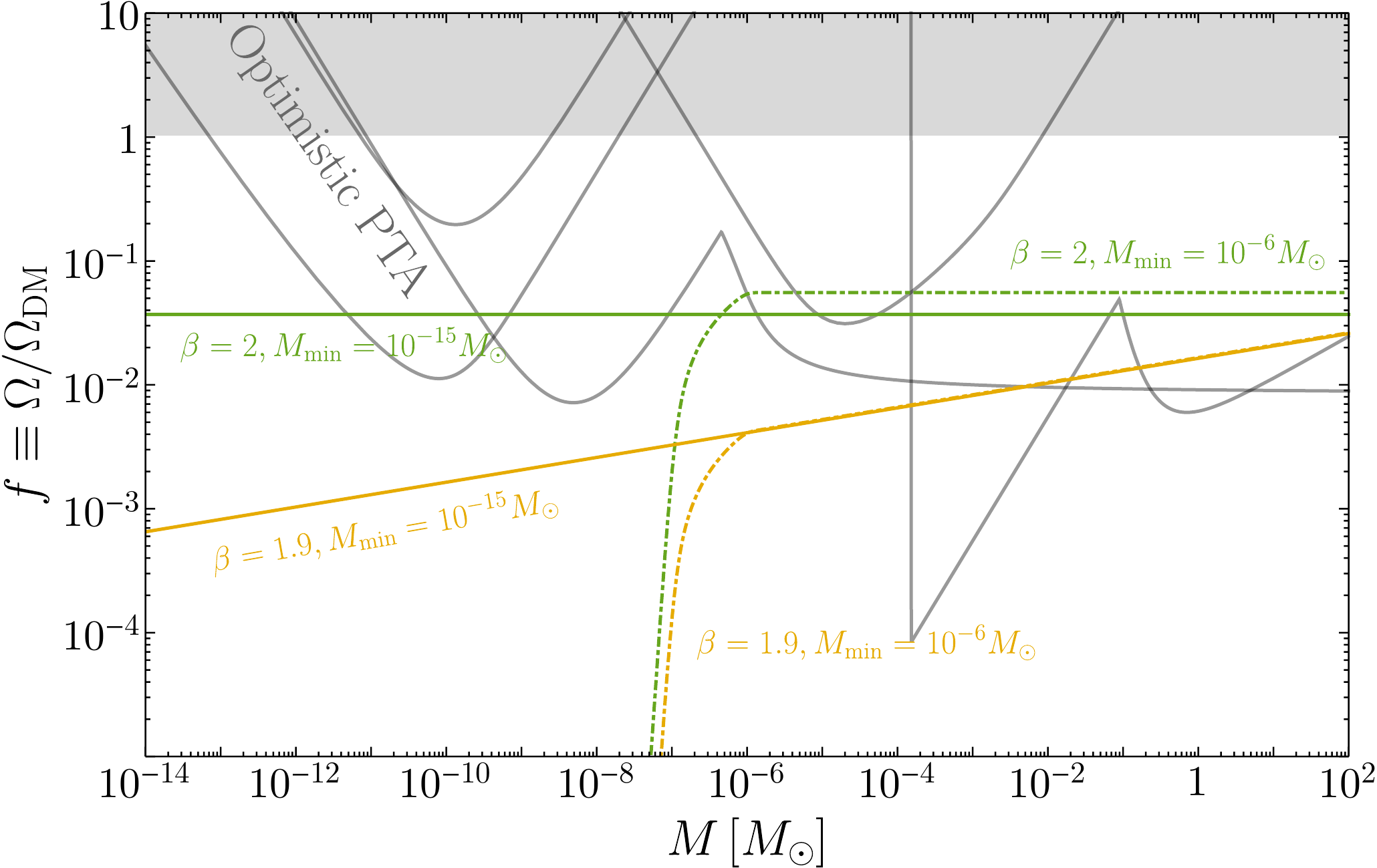}
\caption{Comparison of the mass fraction of dark matter in $[M, 10M]$, $\int_M^{10M} (F(M')/\rho_\text{DM}) \, dM'$, and monochromatic PBH constraints (gray) from Fig.~\eqref{fig:monochromatic_mass_constraints_opt}. The Halo Mass Function is labeled by a spectral index $\beta$ and minimum subhalo mass $M_{\rm min}$ where the drop-off in $M_\text{min}$ has been smoothed from Eq.~\eqref{eq:CDMHMF_norm}. The values of $\beta$ shown are motivated by scale invariant density perturbations seeded during inflation ({\em i.e.} CDM HMF). Comparing the gray PTA reach curves against the prevalence of subhalos in a particular mass bin (green or yellow curves) roughly shows for which mass subhalos PTA reach, via the procedure summarized in Eq.~\eqref{eq:carr}, is viable.}
\label{fig:pbh_hmf_overlay}
\end{figure}

It is clear from Eq.~\eqref{eq:carr} that there is enhanced sensitivity to Halo Mass Functions $F(M)$ with large support to masses for which a particular type of PTA search is sensitive. We show this in Fig.~(\ref{fig:pbh_hmf_overlay}) by comparing PTA search constraints from Fig.~(\ref{fig:monochromatic_mass_constraints_opt}) and the mass fraction in $[M, 10M]$, $\int_M^{10M} (F(M')/\rho_\text{DM}) \, dM'$, for a few different HMFs.  
This translates to sensitivity in Fig.~(\ref{fig:extendedHMF}) when $M_\min$ and $M_\max$ encapsulate the entire mass sensitivity range of a particular type of search. In addition, if $M_\max - M_\min$ is substantially larger than the sensitivity range of a particular search, this results in the reduction of reach to such dark matter masses; this is simply because the fraction of dark matter in the sensitivity range is diluted. Across the board, this is seen in weaker limits for $M_{\max}=10^{12}~M_\odot$ compared to $M_{\max}=10^8~M_\odot$. However this reduction scales only as $f\sim 1/\log\left(M_\max/M_\min\right)$ for $\beta=2$, such that there is only a logarithmic decline for small enough $M_\min$ in the top row, left and middle panels of Fig.~(\ref{fig:extendedHMF}). For $\beta=1.9$, $f\sim 1/M^{0.1}_\max$ so that the reach curves flatten out for small enough $M_\min$ in the left and middle panels. 

For $M_\max=10^{12}~M_\odot$ and $M_\max=10^8~M_\odot$ with $\beta=2$, non-trivial constraints will be set with an SKA-like PTA for $c=10^4$ and above, from the deterministic Doppler and Shapiro constraints derived in Ref.~\cite{Dror2019a}, primarily due to the sensitivity to a wide range in mass. There is less reach for $\beta=1.9$ due to the skew of the HMF towards larger mass subhalos, as can be seen in Fig.~(\ref{fig:pbh_hmf_overlay}). There are more optimistic projections when $M_\max =10^3 M_\min$, even for small concentration parameters.

In order to demonstrate what will be necessary from PTAs to probe CDM-like substructure, we also show improvements in projected reach as some PTA parameters are dialed to more optimistic values in Fig.~(\ref{fig:extendedHMF_optimistic}). We take the SKA-like parameters discussed in Sec.~\ref{subsec:signalSNR} and increase the observing time to $30$ years in the left panel,  the number of pulsars to $1000$ in the middle panel and finally a combined set of optimistic parameters ($N_P = 1000$, $T = 30 \text{ yr}$, $t_\text{rms} = 10 \text{ ns}$, $\Delta t = 1 \text{ week}$, $z_0 = 10 \text{ kpc}$) in the right panel.  $30$ years of observation time will allow reach to HMFs with $\beta = 2$ and subhalos with $c=100$ via the deterministic event Doppler constraints from Ref.~\cite{Dror2019a}. With the optimistic set of PTA parameters shown in the right-hand panel, the stochastic Doppler signal can reach subhalos with concentration parameters $c=10$ composing less than two percent of dark matter. 

\subsection{Backgrounds}

Similar to direct detection of dark matter, non-observation of a signal allows one to set constraints, but a claim for discovery requires careful noise discrimination. While the noise modeling adopted in this work assumes only white noise, red-noise has been observed in some pulsars primarily due to intrinsic deviations from the spin-down timing model \cite{Wang:2015bsa} and fluctuations in the dispersion measure \cite{Becker:2017yyc}. Similar to stochastic gravitational waves, the signal can be differentiated from the background exploiting the differences in the power spectral indices \cite{Wang:2015bsa}. The contributions from the interstellar medium are dependent on the pulsar light frequency and could be corrected for without removing the pulsar-frequency-independent dark matter signal. Furthermore in the case of the Earth term, variations unique to individual pulsars are suppressed when considering cross-correlations between pulsars. 
      
Gravitational waves from supermassive black hole merger events are expected to be detected before a dark matter signal and hence will constitute another background to a dark matter signal \cite{lommen2015pulsar}. A single merger event can be differentiated from the deterministic dark matter subhalo event via the characteristic signal shape, as outlined in Ref.~\cite{Dror2019a}. The stochastic gravitational wave background will also have a different power spectral index compared to the stochastic dark matter signal analyzed in this work. Furthermore, dark matter signals are dipolar in nature and exhibit characteristically different angular correlations compared to gravitational waves which are quadrupolar signal. See {\em e.g.} Ref.~\cite{Jenet:2014bea} for the analog of the Hellings and Downs analysis \cite{Hellings:1983fr} for other signal patterns.
   
Finally, baryonic objects could cause Doppler and Shapiro delays identical to dark matter subhalos in the mass range of sensitivity. However the baryonic matter is dominated by stars in the solar mass range \cite{Fukugita:2004ee}, which would only affect the static signals. A discovery here would necessarily involve supplemental analysis with luminosity discrimination. At lower masses ($10^{-2}-10^{-1} M_\odot$), sub-stellar objects including brown dwarfs make up less than one percent of the total baryon density. At even lower masses, planets make up less than $0.005\%$ of all the baryons. Furthermore these objects have to be transiting and not  bound to the pulsar/solar system in order to mimic a dark matter signal. Nonetheless, pulsars in baryon rich environments might very well display irreducible backgrounds from these loose baryonic objects in certain mass ranges. However, for high latitude pulsars which have been discovered recently~\cite{lorimer2008binary}, the baryonic background will not be limiting in the near future due to the strong limits set on by gravitational MACHO lensing surveys. 

\section{Conclusions}
\label{sec:conclusion}

Pulsar Timing Arrays offer a unique glimpse of dark matter substructure at previously inaccessible small scales.  Current constraints from measurements of large scale structure, for diffuse CDM-like subhalos, become weak below $\sim 10^7~M_\odot$.  For objects as dense as primordial black holes, observational limits from disruption of structure in conjunction with lensing extend the reach down to $\sim 10^{-10}~M_\odot$, though these limits are rapidly lifted as soon as the subhalos become become even modestly less dense.  By contrast, the methods presented here, and in our companion paper \cite{Dror2019a}, reach subhalos having a concentration parameter as small as $c = 10$ and as light as $\sim 10^{-13}~M_\odot$. PTAs are thus potentially more powerful than other existing or proposed probes of dark matter substructure.

As summarized in Figs.~(\ref{fig:monochromatic_mass_constraints}),~(\ref{fig:monochromatic_mass_constraints_opt}), we showed that future PTAs will be able to place strong constraints on the fraction of dark matter in such substructure, even for NFW subhalos having a concentration as small as $c \sim 10$ where lensing cannot reach. We also found that our analysis utilizing multiple transiting subhalos causing a stochastic signal can extend constraints for both Doppler and Shapiro signals by more than an order of magnitude to smaller mass relative to the constraint obtained from a deterministic event of a single transiting subhalo~\cite{Dror2019a}, as seen by comparing the curves labeled `stoch' to the other curves in Fig.~(\ref{fig:monochromatic_mass_constraints}).

A definitive goal is to observe substructure consistent with ordinary CDM at as small of a mass scale as possible, to see to what mass scale such substructures persist.  For the SKA-like PTA parameters described at the end of Sec.~\ref{subsec:signalSNR}, we find this will be difficult unless the power spectrum is significantly more skewed towards low mass subhalos than we expect from standard CDM, as shown in Fig.~(\ref{fig:extendedHMF}).  We showed the requirements from a PTA in Fig.~(\ref{fig:extendedHMF_optimistic}) to reach standard CDM.  While in principle possible, standard CDM offers a challenge, though one that would offer an unprecedented view if reached.  

On the other hand, dark matter often has dynamics that, on small scales, enhance the density perturbations seeded by adiabatic, scale-invariant inflation.  The formalism developed here is sufficiently general to account for any HMF. In future work we will apply this to other well motivated forms of substructure, such as axion miniclusters, cosmologies with a period of matter domination, and vector bosons produced during inflation.  Observing dark matter substructure on small scales gives a unique window into post-inflationary dynamics and the nature of the dark matter.

\acknowledgments

We thank Nikita Blinov and Jeff Dror for useful discussions, and Steve Taylor and Michele Vallisneri for discussions on the NANOGrav data and analysis.  H.R. is supported in part by the DOE under contract DE-AC02-05CH11231. Some of this work was done at the Aspen Center for Physics, which is supported by NSF grant PHY-1607611 and at KITP, supported in part by the National Science Foundation under Grant No. NSF PHY-1748958. T.T. would like to thank the Walter Burke Institute for Theoretical Physics for hospitality during the completion of this work.

\newpage
\appendix

\section{Derivation of the Optimal SNR}
\label{app:SNR_derivation}

We derive the optimal signal-to-noise ratio (SNR) for different signals using a matched filter procedure, similar to Refs.~\cite{Smith2019a,Moore2015a,Allen1999a}. To generalize our discussion we define the measured signal, $s_I(t)$, the dark matter signal $h_I(t)$, and the noise in the detector, $n_I(t)$, which satisfies the relation $s_I = h_I + n_I$ for pulsar $I$. We will derive the optimal SNR for a deterministic signal, where $h_I$ is known, as well as a stochastic signal where $\langle h_I(t) h_J(t') \rangle \equiv R_{IJ}(t, t')$ is known. We will consider both a \textit{pulsar} term, which assumes an independent signal in each pulsar, and an \textit{Earth} term, where correlations between pulsars can boost the SNR. For most of the derivation we assume that the detector noise is stationary and independent across pulsars,
\begin{align}
\langle n_I(t) n_J(t') \rangle = \delta_{IJ} N_I(t - t')\, .
\end{align}
Finally we will simplify the SNR in the limit where the timing residual noise is white and identical in each pulsar: $\langle \delta t_I(t) \delta t_J(t')\rangle = \delta_{IJ} t_\text{rms}^2 \Delta t \, \delta(t - t')$, where $\Delta t$ is the measurement cadence. The residual noise is related to the timing residual by a factor of the pulsar frequency, $n_I = \nu_I \delta t_I$ and therefore\footnote{The signals discussed in the main text also have an accompanying factor of $\nu_I$, and, as we will show, the SNR is independent of this factor.}
\begin{align}
N_I(t - t') & = \nu_I^2 t_\text{rms}^2 \Delta t \, \delta(t - t') \, \\
\widetilde{N}_I & = \nu_I^2 t_\text{rms}^2 \Delta t \, .
\end{align}

\subsection{Deterministic Signal SNR}
\label{subapp:deterministic_snr}

We begin with computing the optimal SNR for a deterministic signal. This derivation will closely follow the discussion given in Ref.~\cite{Moore2015a}. We begin by defining a test statistic, 
\begin{align}
\mathcal{T} = \sum_{I = 1}^{N_P} \int dt \, s_I(t) \, Q_I(t) \, ,
\end{align}
where $Q_I(t)$ is a filter function chosen to maximize the SNR,
\begin{align}
\text{SNR}^2 = \frac{\langle \mathcal{T} \rangle_{s = h + n}^2}{\langle \mathcal{T}^2 \rangle_{s = n} - \langle \mathcal{T} \rangle_{s = n}^2} \, .
\end{align}
The subscripts on $\langle \rangle$ indicate what $s$ is assumed to be. We can compute the expectation values,
\begin{align}
\langle \mathcal{T} \rangle_{s=n} & = 0 \\
\langle \mathcal{T} \rangle_{s = h + n} & = \sum_{I = 1}^{N_P} \int dt \, h_I(t) Q_I(t) \nonumber \\
& = \sum_{I = 1}^{N_P} \int dt d\mathfrak{f}_1 d\mathfrak{f}_2 \, e^{2 \pi i (\mathfrak{f}_1 + \mathfrak{f}_2) t} \, \widetilde{h}_I(\mathfrak{f}_1) \widetilde{Q}_I(\mathfrak{f}_2) \nonumber \\
& =  \sum_{I = 1}^{N_P} \int d\mathfrak{f} \, \widetilde{h}_I(\mathfrak{f}) \widetilde{Q}_I^*(\mathfrak{f}) \\
\langle \mathcal{T}^2 \rangle_{s = n} & =  \sum_{I = 1}^{N_P} \sum_{J = 1}^{N_P} \int dt dt' \, \langle n_I(t) n_J(t') \rangle Q_I(t) Q_J(t') \nonumber \\
& = \sum_{I = 1}^{N_P} \int dt dt' \, N_I(t - t') Q_I(t) Q_I(t') \nonumber \\
& =  \sum_{I = 1}^{N_P} \int dt dt' d\mathfrak{f}_1 d\mathfrak{f}_2 d\mathfrak{f}_3 \,  e^{2 \pi i (\mathfrak{f}_1 + \mathfrak{f}_2) t} e^{2 \pi i (\mathfrak{f}_3 - \mathfrak{f}_1) t'} \widetilde{N}_I(\mathfrak{f}_1) \widetilde{Q}_I(\mathfrak{f}_2) \widetilde{Q}_I(\mathfrak{f}_3) \nonumber \\
& = \sum_{I = 1}^{N_P} \int d\mathfrak{f} \, \widetilde{N}_I(\mathfrak{f}) \widetilde{Q}_I^*(\mathfrak{f}) \widetilde{Q}_I(\mathfrak{f}) \, ,
\end{align}
The SNR is then,
\begin{align}
\text{SNR}^2 = \frac{\left| \sum\limits_{I = 1}^{N_P} \displaystyle\int d\mathfrak{f} \, \widetilde{h}_I(\mathfrak{f}) \widetilde{Q}_I^*(\mathfrak{f}) \right|^2}{\sum\limits_{I = 1}^{N_P} \displaystyle\int d\mathfrak{f} \, \widetilde{N}_I(\mathfrak{f}) \widetilde{Q}_I^*(\mathfrak{f}) \widetilde{Q}_I(\mathfrak{f}) } \, .
\end{align}
The $Q(\mathfrak{f})$ which optimizes this SNR is $Q_I(\mathfrak{f}) = \widetilde{h}_I(\mathfrak{f})/\widetilde{N}_I$, and therefore the optimal SNR is
\begin{align}
\text{SNR}^2 = \sum\limits_{I = 1}^{N_P} \int d\mathfrak{f} \, \frac{\left| \widetilde{h}_I(\mathfrak{f}) \right|^2}{\widetilde{N}_I(\mathfrak{f})} \, .
\end{align}
Assuming the noise is white and pulsar independent we can further simplify,
\begin{align}
\text{SNR}^2 = \frac{1}{\widetilde{N}} \sum_{I = 1}^{N_P} \int dt \, h_I^2(t) \, .
\label{eq:deterministic_snr}
\end{align}
If the signal is independent in each pulsar and the SNR is dominated by the largest signal across the array, then the SNR can be approximated as
\begin{align}
\text{SNR}^2 = \frac{1}{\widetilde{N}} \, \underset{\{I\}}{\max} \left[ \int dt \,  h_I^2(t) \right] \, ,
\end{align}
which we define as the \textit{pulsar} term SNR. If the signal has a similar amplitude, but not independent across the pulsars, then Eq.~\eqref{eq:deterministic_snr} will be parametrically larger by a factor of $N_P$. We define this as the \textit{Earth} term SNR, because it's the SNR used when the dark matter interacts with the Earth. We can simplify this further by replacing the sum with an average over the pulsar positions, defined by $\langle \rangle_\mathcal{P}$
\begin{align}
\text{SNR}^2 = \frac{N_P}{\widetilde{N}}\int dt \, \langle h_I^2(t) \rangle_\mathcal{P} \, .
\end{align}

\subsection{Stochastic Pulsar Term SNR}
\label{subapp:stochastic_snr_pulsar}

We begin by defining a test statistic, $\mathcal{T}$,
\begin{align}
\mathcal{T} = \sum_{I = 1}^{N_P} \int dt dt' \, \left( s_I(t)s_I(t') - \langle n_I(t) n_I(t') \rangle \right) Q_I(t,t')
\label{eq:pulsar_test_statistic}
\end{align}
where $Q_I(t, t')$ is a filter function applied to the time series of the $I$\th pulsar. Our goal is to find the $Q_I$'s which maximize the SNR,  
\begin{align}
\text{SNR}^2 = \frac{\langle \mathcal{T} \rangle_{s = h + n}^2}{ \langle \mathcal{T}^2 \rangle_{s = n} - \langle \mathcal{T} \rangle_{s = n}^2 },
\label{eq:SNR_definition}
\end{align}
where the subscripts denote the assumptions under which we should evaluate the expectation values. We have
\begin{align}
\langle \mathcal{T} \rangle_{s = n} & = 0 \\
\langle \mathcal{T} \rangle_{s = h+n} & = \sum_{I = 1}^{N_P} \int dt dt' \langle h_I(t) h_I(t') \rangle Q_I(t, t') \nonumber \\
& = \sum_{I = 1}^{N_P} \int dt dt' \prod_{k=1}^4 df_k \,  e^{2 \pi i (\mathfrak{f}_1 + \mathfrak{f}_3) t} e^{2 \pi i (\mathfrak{f}_2 + \mathfrak{f}_4) t'} \langle \widetilde{h}_I(\mathfrak{f}_1) \widetilde{h}_I(\mathfrak{f}_2) \rangle \widetilde{Q}_I(\mathfrak{f}_3, \mathfrak{f}_4) \nonumber \\
& = \sum_{i = 1}^{N_P} \int d\mathfrak{f} d\mathfrak{f}' \, S_I(\mathfrak{f}, \mathfrak{f}') \widetilde{Q}_I^*(\mathfrak{f}, \mathfrak{f}') \\
\langle \mathcal{T}^2 \rangle_{s = n} & = \sum_{I=1}^{N_P} \sum_{J = 1}^{N_P} \int \prod_{k=1}^4 dt_k \, \left( \langle n_I(t_1) n_I(t_2) n_J(t_3) n_J(t_4) \rangle - N_I(t_1 - t_2) N_J(t_3 - t_4) \right) \nonumber\\
& \quad \quad \quad \quad\quad\quad \times Q_I(t_1, t_2) Q_J(t_3, t_4) \, ,
\label{eq:noise_T2}
\end{align}
where $S_I(\mathfrak{f}, \mathfrak{f}') \equiv \langle \widetilde{h}_I(\mathfrak{f}) \widetilde{h}_I(\mathfrak{f}') \rangle$. Evaluating the four point function of the noise, 
\begin{align}
\langle n_I(t_1) n_I(t_2) n_J(t_3) n_J(t_4) \rangle & = N_I(t_1 - t_2) N_J(t_3 - t_4) \nonumber \\
& + \delta_{IJ} \left( N_I(t_1 - t_3) N_J(t_2 - t_4) + N_I(t_1 - t_4) N_J(t_2 - t_3) \right) \, ,
\end{align}
which allows us to simplify Eq.~\eqref{eq:noise_T2}. The two remaining terms are identical, and the whole expression can be simplified to
\begin{align}
\langle \mathcal{T}^2 \rangle_{s = n} & = 2 \sum_{I = 1}^{N_P} \int d\mathfrak{f} d\mathfrak{f}' \,  \widetilde{N}_I(\mathfrak{f}) \widetilde{N}_I(\mathfrak{f}') \widetilde{Q}_I(\mathfrak{f}) \widetilde{Q}_I^*(\mathfrak{f}').  
\end{align}
The $\widetilde{Q}$ which maximizes the SNR is $Q_I(\mathfrak{f}, \mathfrak{f}') = S_I(\mathfrak{f}, \mathfrak{f}') / ( \widetilde{N}_I(\mathfrak{f}) \widetilde{N}_I(\mathfrak{f}') )$ and therefore the optimal SNR is
\begin{align}
\text{SNR}^2 = \frac{1}{2} \sum_{I = 1}^{N_P} \int d\mathfrak{f} d\mathfrak{f}' \, \frac{|S_I(\mathfrak{f}, \mathfrak{f}')|^2}{\widetilde{N}_I(\mathfrak{f}) \widetilde{N}_I(\mathfrak{f}')} \, .
\end{align}
We can simplify further by assuming that the signal and noise are independent of the pulsar, and that the noise is white,
\begin{align}
\text{SNR}^2 = \frac{N_P}{2 \widetilde{N}^2} \int dt dt' R(t, t')^2 \, .
\end{align}

\subsection{Stochastic Earth Term SNR}
\label{subapp:stochastic_snr_Earth}

The derivation of the optimal Earth term SNR is similar to the pulsar term, except the test statistic is slightly different,
\begin{align}
\mathcal{T} = \sum_{I \neq J}^{N_P(N_P - 1)} \int dt dt' s_I(t) s_J(t') Q_{IJ}(t, t'),
\end{align}
where the sum is over pairs of pulsars. There is no subtracted piece, as there is in Eq.~\eqref{eq:pulsar_test_statistic}, because the second term evaluates to zero when $I \neq J$. Computing the terms in the SNR gives, 
\begin{align}
\langle \mathcal{T} \rangle_{s = n} & = 0 \\
\langle \mathcal{T} \rangle_{s = h + n} & = \sum_{I \neq J} \int dt dt' \, \langle h_I(t) h_J(t') \rangle Q_{IJ}(t, t') \nonumber \\
\langle \mathcal{T} \rangle_{s = h + n} & = \sum_{I \neq J} \int d\mathfrak{f} d\mathfrak{f}' \, S_{IJ}(\mathfrak{f}, \mathfrak{f}') \widetilde{Q}^*_{IJ}(\mathfrak{f}, \mathfrak{f}') \\
\langle \mathcal{T}^2 \rangle_{s = n} & = \sum_{I \neq J} \sum_{K \neq L} \int \prod_{k=1}^4 dt_k \, \langle n_I(t_1) n_J(t_2) n_K(t_3) n_L(t_4)\rangle Q_{IJ}(t_1, t_2) Q_{KL}(t_3, t_4) \, ,
\end{align}
where $S_{IJ}(\mathfrak{f}, \mathfrak{f}') \equiv \langle \widetilde{h}_I(\mathfrak{f}) \widetilde{h}_J(\mathfrak{f}') \rangle$ and we use a finite time delta function to remove the $t$ integrals. Computing the four point function, noting that $I \neq J$, $K \neq L$ by definition gives,
\begin{align}
\langle n_I(t_1) n_J(t_2) n_K(t_3) n_L(t_4) \rangle = \delta_{IK} \delta_{JL} N_{I}(t_1 - t_3) N_{J}(t_2 - t_4) + \delta_{IL} \delta_{JK} N_{I}(t_1 - t_4) N_{J}(t_2 - t_3) \, .
\end{align}
Substituting this expression in $\langle \mathcal{T}^2 \rangle_{s=n}$ gives two identical terms which we can simplify to
\begin{align}
\langle \mathcal{T}^2 \rangle_{s = n} & = 2 \sum_{I \neq J} \int d\mathfrak{f} d\mathfrak{f}' \widetilde{N}_I(\mathfrak{f}) \widetilde{N}_J(\mathfrak{f}') \widetilde{Q}_{IJ}^*(\mathfrak{f}) \widetilde{Q}_{IJ}(\mathfrak{f}') \, .
\end{align}
Again one can show that the optimal $\widetilde{Q}_{IJ}(\mathfrak{f}, \mathfrak{f}')$ is $S_{IJ}(\mathfrak{f}, \mathfrak{f}')/ \left( \widetilde{N}_I(\mathfrak{f}) \widetilde{N}_J(\mathfrak{f}') \right)$ and therefore the optimal SNR is given by
\begin{align}
\text{SNR}^2 = \frac{1}{2} \sum_{I \neq J} \int d\mathfrak{f} d\mathfrak{f}' \frac{\left| S_{IJ}(\mathfrak{f}, \mathfrak{f}') \right|^2}{\widetilde{N}_I(\mathfrak{f}) \widetilde{N}_J(\mathfrak{f}')} \, .
\end{align}
Lastly, we assume that the noise is white and identical across pulsars, and replace the sum by taking an average over the pulsar positions, denoted by $\langle \rangle_{\mathcal{P}}$,
\begin{align}
\text{SNR}^2 = \frac{N_P(N_P - 1)}{2 \widetilde{N}^2} \int dt dt' \left\langle R_{IJ}(t, t')^2 \right\rangle_{\mathcal{P}} 
\end{align}

\section{Subtraction of Best Fit Parameters in PTA Signal}
\label{app:subtraction_procedure}

\begin{figure}[ht]
\includegraphics[width=\textwidth]{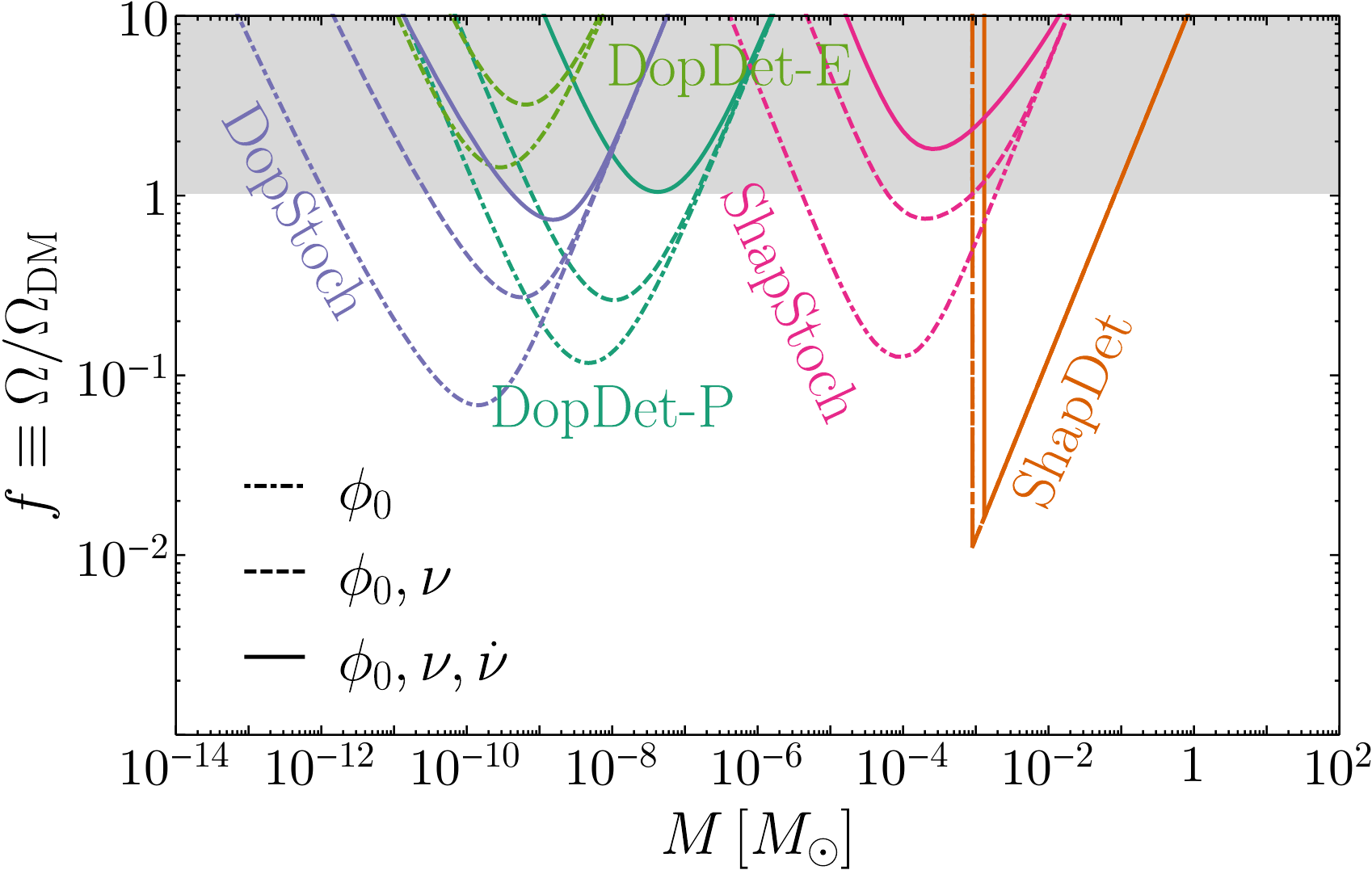}
\caption{Constraints of the fraction of dark matter, $f = \Omega / \Omega_\dm$ in monochromatic PBHs (similar to Fig.~(\ref{fig:monochromatic_mass_constraints})) when different number of parameters are included in the timing model for deterministic and stochastic constraints. Static constraints are not shown since the fitting procedure trivially picks out only the second derivative.}
\label{fig:subtraction_constraints}
\end{figure}

We discuss how the parameters of the pulsar timing model, \textit{e.g.} $\phi^0_\fit, \nu_\fit, \dot{\nu}_\fit, ...$, impacts the inferred dark matter signal. Let $\phi_m(t)$ denote the phase that is measured at time $t$, and therefore the goodness of fit is characterized by \footnote{In a PTA measurement, $\phi_M$ is measured at an arrival time of the $n$\th pulse such that $\phi(t_n)$ is subtracted in the standard discrete time formulation of $\chi^2$. The analysis here only differs by working in continuous time.}
\begin{align}
\chi^2 = \frac{1}{T} \int \left( \phi_m(t) - \phi^0 - \nu t - \frac{1}{2} \dot{\nu}t^2 \right)^2 dt \, . \label{eq:chisq_app}
\end{align}
The generalization to a timing model with higher order terms is straightforward. In order to find the best-fit parameters we minimize $\chi^2$ with respect to $\phi^0, \nu, \dot{\nu}$, which is more easily done by defining an inner product, 
\begin{align}
\left( a, b \right) \equiv \frac{1}{T} \int a(t) b(t)\,  dt \, ,
\end{align} 
along with a set of polynomial basis functions, $f_i$, with respect to this inner product, \footnote{These are related to the standard Legendre polynomials, $P_n$, by a scaling and shift: $f_n(t) = \sqrt{2 n + 1} \, P_n(2 t/T - 1)$.}
\begin{align}
\left( f_i, f_j \right) & = \delta_{ij} \, .
\end{align}
Therefore $\chi^2$ in Eq.~\eqref{eq:chisq_app} can be written as
\begin{align}
\chi^2 = \left( \phi_m(t) - \sum_{k=0}^{2} c_k f_k(t), \phi_m(t) - \sum_{k=0}^2 c_k f_k(t)  \right),
\end{align}
where $c_k = \left(\phi_0 + \nu t + \dot{\nu} t^2/2, f_k \right)$. We can now minimize with respect to $c_k$ (and later work out $\phi^0_\fit, \nu_\fit, \dot{\nu}_\fit$ if necessary). The minimization condition is
\begin{align}
\frac{\partial \chi^2}{\partial c_k}(c_k^\fit) & = -2 \left( f_k, \phi_m \right) + 2 \sum_l c_l^\fit \left( f_k, f_l  \right) = -2 \left( f_k, \phi_m \right) + 2 c_k^\fit = 0 \, ,
\end{align}
and therefore $c_k^\fit = \left( \phi_m, f_k \right)$. The residual, $s$, is then given by $s = \phi_m(t) - \sum_k c_k^\fit f_k(t)$, and the subtracted DM signal, $h$, is related to the unsubtracted signal, $\delta \phi$ by,
\begin{align}
h(t) = \delta \phi(t) - \sum_{k = 0}^2 (\delta \phi, f_k) f_k(t) \, .
\label{eq:single_subtracted_signal}
\end{align}
This is how the subtraction procedure effects the single deterministic event analysis, as the SNR is only a function of $h$. However for the stochastic signal the subtraction procedure enters through a correlator of subtracted signals. We can relate the unsubtracted correlator, $R$, to the subtracted correlator, $R_\text{sub}$, by
\begin{align}
R_\text{sub}(t, t') & \equiv \langle h(t) h(t') \rangle \nonumber  \\
& = \left\langle \left(\delta \phi(t) - \sum_{n = 0}^2 \left( \delta \phi, f_n \right) f_n(t)  \right) \left( \delta \phi(t') - \sum_{n = 0}^2 \left( \delta \phi, f_n \right) f_n(t')  \right)  \right\rangle \nonumber \\
& = R(t, t') - \sum_{n = 0}^2 f_n(t) \mathcal{R}_n(t') - \sum_{n = 0}^2 f_n(t') \mathcal{R}_n(t) + \sum_{n = 0}^2 \sum_{m=0}^2  f_n(t) f_m(t') \mathcal{R}_{nm} \, ,
\label{eq:sub_auto_corr}
\end{align}
where
\begin{align}
\mathcal{R}_n(t') & \equiv \left( R(t, t'), f_n(t) \right) \\
\mathcal{R}_{nm} & \equiv \frac{1}{T^2} \int dt dt' \, R(t, t') f_n(t) f_m(t') \, .
\end{align}
The results of this procedure are quoted in Eq.~\eqref{eq:sub_auto_corr_main}.  The effect of this subtraction procedure on the reach to monochromatic PBHs can be seen in Fig.~(\ref{fig:subtraction_constraints}).  Note that in this figure we have not included `Static' constraints, as these are derived with only $\ddot \nu$ and higher order terms, such that subtraction cannot be meaningfully applied to this analysis.

\section{Relationship with the Power Spectrum Approach}
\label{powerallspace}
In this section, we explain the difference between this work and the power spectrum procedure adopted in Ref.~\cite{Baghram2011a}. 
If the observing volume is all of space then $R$ in Eq.~\eqref{eq:R1_correlator}, for the Doppler and Shapiro delays, can be written in terms of the matter power spectrum, as done in Ref.~\cite{Baghram2011a}.\footnote{The $\langle \delta \phi(t) \delta \phi(t') \rangle$ signal correlator we consider here and $\left\langle \frac{\delta \nu}{\nu}(t) \frac{\delta \nu}{\nu}(t') \right\rangle$ considered in Ref.~\cite{Baghram2011a} are related by $\langle \delta \phi(t) \delta \phi(t') \rangle = \nu^2 \int_0^t \int_0^{t'} dt_1 dt_2 \left\langle \frac{\delta \nu}{\nu}(t_1) \frac{\delta \nu}{\nu}(t_2)  \right\rangle$.} To show this, we simplify the gravitational potential correlator, from which both the Doppler and Shapiro signal correlators can be derived. For example, the Shapiro delay frequency shift correlator is
\begin{align}
\left\langle \frac{\delta \nu}{\nu}(t) \frac{\delta \nu}{\nu}(t') \right\rangle = 4  \frac{d}{dt} \frac{d}{dt'} \int_0^{z_0} \int_0^{z_0} dz_1 dz_2 \int \frac{d^3\vec{k}d^3\vec{k}'}{(2\pi)^6} e^{i \left( \vec{k} + \vec{k}' \right) \cdot \vec{r} } \langle \widetilde{\Phi}(\vec{k}, t; Q) \widetilde{\Phi}(\vec{k}', t'; Q) \rangle 
\end{align}
where $\widetilde{\Phi}(\vec{k}; Q)$ is the Fourier transform of the gravitational potential at position $\vec{r}$, and $Q$ represents all of the random variables, $\vec{r}^0_1, \vec{r}^0_2, M_1, M_2,$ etc.
 
Since the object's individual potentials only depend on the distance from the center of mass, $\vec{r}_i(t) = \vec{v}t + \vec{r}^0_i$,
\begin{align}
\widetilde{\Phi}(\vec{k}, t; Q) = \sum_i e^{i \vec{k} \cdot \vec{r}_i(t)} \widetilde{\Phi}_i(\vec{k}; M_i) \, .
\end{align} 
The potential correlator is then a sum over individual contributions as,
\begin{align}
\langle \widetilde{\Phi}(\vec{k}, t; Q) \widetilde{\Phi}(\vec{k}', t'; Q) \rangle = \sum_{i, j} \left\langle e^{i \left(\vec{k} \cdot \vec{r}_i + \vec{k}' \cdot \vec{r}_j \right) } \right\rangle_{\vec{r}^0} \langle \widetilde{\Phi}_i(\vec{k}, M_i) \widetilde{\Phi}_j(\vec{k}', M_j) \rangle_M \, , \label{eq:phi_corr_sum}
\end{align}
where the subscripted $\langle \rangle$ denotes averaging over only the subscripted random variable. In the limit where the $\vec{r}^0$ integral is over all space, the exponential in the $i = j$ term becomes a delta function, and the potential correlator from a single subhalo can be simplified,
\begin{align}
\left\langle e^{i \left(\vec{k} \cdot \vec{r}^0_i + \vec{k}' \cdot \vec{r}^0_i \right) } \right\rangle_{\vec{r}^0} & = \frac{(2\pi)^3}{V} \delta^{3}(\vec{k} + \vec{k}') \\
\langle \widetilde{\Phi}_i(\vec{k}, M_i) \widetilde{\Phi}_i(\vec{k}, M_i) \rangle_M &  = \frac{16 \pi^2 G^2 \overline{n}^2}{k^4} P_m^{1h}(k) \, ,
\end{align}
where $P^{1h}_m$ is the 1-subhalo matter power spectrum. It can be shown that the $i \neq j$ term in Eq.~\eqref{eq:phi_corr_sum} simplifies similarly with $P_m^{1h}$ being replaced by $P_m^{2h}$. Finally,
\begin{align}
\langle \widetilde{\Phi}(\vec{k}, t; Q) \widetilde{\Phi}(\vec{k}', t'; Q) \rangle = (2 \pi)^3 \delta^3\left( \vec{k} + \vec{k}' \right)\frac{16 \pi^2 G^2 \overline{n}^2}{k^4} e^{i \vec{k} \cdot \vec{v} (t - t')} \left( P^{1h}_m(k) + P^{2h}_m(k) \right) \, ,
\end{align}
and the rest of the derivation for the signal correlator proceeds as in Ref.~\cite{Baghram2011a}. However the Earth-pulsar system samples a finite volume of space, since subhalos at a larger distance are static, and hence susceptible to being absorbed in the fit. To obtain a physical result from a power spectrum approach one would need to introduce cuts on the $\vec{k}$ integral to incorporate these finite volume effects. Here we simplify the signal correlators differently than Ref.~\cite{Baghram2011a} by incorporating finite volume effects directly in the subhalo position average.
\section{\texorpdfstring{$\mathcal{F}$}{TEXT} Form Factor Integral}
\label{app:form_factor}

We derive an analytic expression for the form factor in Eq.~\eqref{eq:F_int},
\begin{align}
\mathcal{F}(M, b) = b \int dk \,  W(k, M) J_1(kb) \, ,
\label{eq:F_integral}
\end{align}
with the definition of $W$,
\begin{align}
W(k, M) = \frac{4 \pi}{M} \int_0^{r_v} r^2 \text{sinc}(k r) \rho(r, M) dr \, .
\end{align}
Substituting this in to Eq.~\eqref{eq:F_integral},
\begin{align}
\mathcal{F}(M, b) & = \frac{4 \pi b}{M} \int_0^{r_v} r^2 \rho(r, M) dr \int dk \,  J_1(kb) \text{sinc}(k r) \, ,
\end{align}
and using
\begin{align}
\int dk J_1(kb) \text{sinc}(k r) = \frac{1}{b} \left(1 - \Theta(r - b) \sqrt{1 - \frac{b^2}{r^2}} \right) \, ,
\end{align}
gives
\begin{align}
\mathcal{F}(M, b) & = 1 - \frac{4 \pi}{M} \int_b^{r_v} \sqrt{1 - \frac{b^2}{r^2}} r^2 \rho(r, M) dr \, .
\end{align}
Assuming $\rho$ is given by an NFW profile then
\begin{align}
\rho & = \rho_s \frac{1}{\left( \frac{r}{r_s} \right) \left(1 + \frac{r}{r_s} \right)^2} \\
\rho_s & = \frac{1}{r_s^3} \frac{M}{4 \pi} \frac{1}{\left( \log{\left(1 + c \right)} - \frac{c}{1 + c} \right)} \, ,
\end{align}
and therefore
\begin{align}
\mathcal{F}(M, b) & = 1 - \frac{1}{r_v^3} \frac{c^3}{\left( \log{\left( 1 + c \right)} - \frac{c}{1 + c} \right)} \int_b^{r_v} \sqrt{1 - \frac{b^2}{r^2}} r^2 \frac{1}{\left( \frac{r}{r_s}\right) \left( 1 + \frac{r}{r_s}\right)^2} dr \, .
\end{align}
We can simplify further by changing variables. Defining $y \equiv r/r_v$, and $x \equiv b/r_v$
\begin{align}
\mathcal{F}\left( x, c \right) & = 1 - \frac{c^2}{\left( \log{\left( 1 + c \right)} - \frac{c}{1 + c} \right)} \int_x^1 \sqrt{1 - \frac{x^2}{y^2}} \frac{y}{\left(1 + cy\right)^2} dy
\end{align}
We see that the form factor $\mathcal{F}$ is only a function of $c$ and $b/r_v$, as shown in Fig.~\ref{fig:formfactor}.

\bibliography{PTA4DM}
\end{document}